\documentstyle[12pt]{article}
\newcommand{\zero}{{(0)}}
\newcommand{\be}{\begin{equation}}
\newcommand{\ee}{\end{equation}}
\newcommand{\ben}{\begin{eqnarray}\displaystyle}
\newcommand{\een}{\end{eqnarray}}
\newcommand{\wh}{\widehat}
\newcommand{\al}{{( a )}}
\newcommand{\alt}{{(a,2)}}
\newcommand{\bet}{{( b )}}
\newcommand{\ten}{{(10)}}
\newcommand{\wc}{\check}
\newcommand{\wt}{\widetilde}

\newcommand{\refb}[1]{(\ref{#1})}
\newcommand{\A}{{\cal A}}
\newcommand{\B}{{\cal B}}
\newcommand{\C}{{\cal C}}
\newcommand{\D}{{\cal D}}
\newcommand{\E}{{\cal E}}
\newcommand{\F}{{\cal F}}
\newcommand{\G}{{\cal G}}
\newcommand{\K}{{\cal K}}
\newcommand{\SS}{{\cal S}}
\newcommand{\T}{{\cal T}}
\newcommand{\LL}{{\cal L}}
\newcommand\M{{\cal M}}
\newcommand{\p}{\partial}
\newcommand{\tF}{\tilde F}

\title{Strong-Weak Coupling Duality in Four Dimensional
String Theory }
\author{Ashoke Sen  \\
Tata Institute of Fundamental Research \\
Homi Bhabha Road, Bombay 400005, INDIA.\\
sen@theory.tifr.res.in, sen@tifrvax.bitnet}

\begin{document}

\maketitle
\smallskip
\begin{abstract}

We present several pieces of evidence for strong-weak coupling duality
symmetry in the heterotic string theory, compactified on a six dimensional
torus.  These include symmetry of the 1) low energy effective action, 2)
allowed spectrum of electric and magnetic charges in the theory, 3)
allowed mass
spectrum of particles saturating the Bogomol'nyi bound, and 4) Yukawa
couplings between massless neutral particles and massive charged particles
saturating the Bogomol'nyi bound.

This duality transformation exchanges the electrically charged elementary
string excitations with the magnetically charged soliton states in the
theory.  It is shown that the existence of a strong-weak coupling duality
symmetry in four dimensional string theory makes definite prediction about
the existence of new stable monopole and dyon states in the theory with
specific degeneracies, including certain supersymmetric bound states of
monopoles and dyons.  The relationship between strong-weak coupling
duality transformation in string theory and target space duality
transformation in the five-brane theory is also discussed.

\end{abstract}

\vfill

{}\vbox{\hbox{TIFR/TH/94-03}
\hbox{hep-th/9402002}}\hfill ~

\eject

\section{Introduction}\label{s1}

String theory has many surprising symmetries which completely
change our understanding of the geometry and topology of
space-time. Among them are the familiar duality symmetries of
string theory compactified on a torus and the mirror symmetries of
string theories compactified on a Calabi-Yau manifold. {}From the
world sheet point of view, these symmetries provide an
equivalence relation between two dimensional quantum field
theories and not between their classical limits. Thus this
equivalence cannot be seen if we expand both the theories in the
$\sigma$-model loop expansion parameter $g_\sigma$, and compare
terms order by order in $g_\sigma$. In this sense, these symmetries
are non-perturbative from the world-sheet point of view.
However, all of these symmetries are valid order by order in
string perturbation theory, {\it i.e.}, the two $\sigma$ models related by
such a symmetry transformation
give rise to equivalent quantum field theories on a two
dimensional surface of any arbitrary genus.

In these notes we shall present evidence that
string theory in four dimensions, resulting from the
compactification of the heterotic string theory on a
six-dimensional torus, possesses another
kind of symmetry, which acts non-trivially on the string loop
expansion parameter $g_{st}$, and hence is not a property of
each term in the expansion in powers of $g_{st}$. In particular,
at the level of states, this duality transformation, acting on
the elementary excitations in string theory carrying electric
charge, gives rise to magnetically charged solitons.  For
definiteness, we shall call this duality transformation
S-duality, and the usual target space duality transformation in
the four dimensional string theory T-duality.  Since at present
the only way we know of calculating anything in string theory is
as a power series expansion in $g_{st}$, we have no way of
actually proving the existence of S-duality symmetry in string
theory. However, there are several quantities in string theory,
where the tree level answers are believed to be the exact
answers. It is possible to check if these quantities are
invariant under the S-duality transformation mentioned above.
We shall focus on four such sets of quantities.

{\bf 1. Low Energy Effective Field Theory}: It is well known that
string theory  at low energies is described by an effective
field theory of masssless fields. {\it A priori} there is no
reason to expect that this field theory will not be modified by
quantum corrections, and in fact, for a generic string
compactification, the low energy effective field theory will be
modified by quantum corrections. However, the theory that we
shall consider, namely the toroidal compactification of the
heterotic string theory, possesses a local $N=4$ supersymmetry in
four dimensions. There is strong evidence that for such
theories, specifying the gauge symmetry group determines the
low energy effective field theory completely\cite{DEROO}. Thus we expect
that the low energy effective
field theory at the tree level is not modified by string quantum
corrections (up to possible redefinitions of various fields).\footnote{We
are implicitly assuming that the computation of the effective action does
not suffer from any infra-red or collinear divergences, so that the
effective action can be expressed as the integral of a local Lagrangian
density.  Since we shall be working at a generic point in the moduli space
of compactification where the unbroken gauge symmetry group is abelian,
and all the charged particles are massive, this is a plausible
assumption.} Thus if S-duality is a genuine symmetry of the theory, this
low energy effective field theory must possess S-duality invariance.

{\bf 2. Allowed Spectrum of Electric and Magnetic Charges}: At a generic
point in the moduli space of vacuum configurations, the theory under
consideration has an unbroken gauge symmetry $U(1)^{28}$.  The U(1)
charges of different states in the theory are described by 28 dimensional
vectors belonging to an even, self-dual, Lorentzian lattice. For N=4
supersymmetric string compactification, we expect
these gauge charges not to be renormalized
by quantum corrections\cite{MARTINEC}.
Since the spectrum of magnetic charges in the
theory is determined from the spectrum of electric charges by the
Dirac-Schwinger-Zwanziger-Witten\cite{DSZ,WITTEN} quantization rules, it
follows that the spectrum of allowed magnetic charges in the theory is
also not renormalized by quantum corrections.  Hence the spectrum of
electric and magnetic charges, calculated from the tree level theory, must
be invariant under the S-duality transformation if it is to be a symmetry
of the theory.

{\bf 3. Allowed Mass Spectrum of Particles
Saturating the Bogomol'nyi Bound}: The
mass of a generic string state is most certainly renormalized by quantum
corrections. However, there is a special class of string states for which
the tree level formul\ae\ for the masses are expected to be
exact\cite{WITTENOLIVE,OSBORN}. These states are characterized by the fact
that 1) they belong to the 16-dimensional representation of the $N=4$
supersymmetry algebra, and 2) their masses are determined completely in
terms of their electric and magnetic charges by the so called Bogomol'nyi
formula, which also gives a lower bound to the mass of any state in the
theory carrying a given amount of electric and magnetic charges. In fact,
the supersymmetry algebra itself constrains the mass of a state in the
16-component supermultiplet to saturate the Bogomol'nyi bound. Since the
representation of a state is not expected to be modified by quantum
corrections, the masses of these states are also expected to be unaffected
by quantum corrections. As a result, if the S-duality transformation is to
be a symmetry of the theory, the allowed mass spectrum of the states in the
16-component supermultiplet, calculated at the tree level, must be
invariant under  this transformation.

{\bf 4. Yukawa Couplings Between Massless Scalars and Massive Charged
States in the 16-component Supermultiplet}: As in the case of the mass
spectrum, the three point couplings between generic string states will
most certainly be modified by quantum corrections. However, as we shall
see, the Yukawa couplings of all the massless scalar fields of the theory
to various string states can be determined in terms of the dependence of
the masses of these states on various modular parameters.  Since we have
already argued that  the masses of the string states belonging to the
16-component supermultiplet are not modified by quantum corrections, the
Yukawa couplings of the massless scalars of the theory to these states
also remain unmodified.  Hence, these Yukawa couplings, calculated at the
tree level, must also remain invariant under the S-duality transformation,
if the latter is a symmetry of the theory.

We shall analyze the S-duality transformation properties of each
of these quantities, and show that they are, indeed, invariant
under this transformation.

S-duality transformation of elementary string states correspond to monopole
and dyon states in the string theory. We shall show that whereas many of
these states can be identified with known monopole and dyon states in the
theory, there are many others which do not correspond to any known state.
Existence of these states can be taken to be a prediction of the S-duality
symmetry.

Besides the conjecture of S-duality symmetry of the four
dimensional string theory, there has been yet another independent
conjecture in string theory which is even harder to test. This
conjecture claims that string theory in ten dimensions is
equivalent to the theory of five-branes (five dimensional
extended objects) in ten dimensions. The reason that this
conjecture is difficult to prove is that 1) the theory of five
branes at lowest order is described by an interacting six
dimensional field theory and has not been solved, and, 2) the
relationship between the loop expansion parameter of the string
theory and that of the five brane theory is somewhat
non-trivial\cite{DUFFLUTWO}, so that the duality conjecture does not
relate a given order term in the string loop expansion parameter to the
same order term in the five-brane loop expansion parameter.

If we accept the equivalence of the string theory and five-brane
theories in ten dimensions despite these difficulties, then it
would also imply the equivalence between the corresponding
theories compactified on a six-dimensional torus. It will then
be natural to ask how the S-duality transformation in string
theory acts on the states of the five-brane theory. It turns out
that the S-duality transformation has a very natural action on the
states of the five-brane theory, namely, it interchanges the
Kaluza-Klein modes of the theory (states carrying non-zero
momenta in the internal direction) with the five-brane winding
modes on the torus. Thus this is an exact analog of the target
space duality (T-duality) transformation in string theory, under
which the Kaluza Klein modes of the theory get exchanged with
the string winding modes on the torus. In this sense, the string
five-brane duality interchanges the roles of the T-duality and
S-duality. We call this `duality of dualities'.

These notes will be divided into two main parts. In the first
part (\S\ref{s2}-\S\ref{snew}) we
shall discuss the evidence for the S-duality symmetry in four
dimensional string theory. In the second part (\S\ref{s4}) we
shall show how the electric-magnetic duality transformation in
string theory can be interpreted as the target space duality
transformation in the five-brane theory compactified on a six
dimensiional torus. Much of the material in these notes will be
a review of Refs.\cite{SEN1,SEN2,SENBOGOM,SCHSEN1,SCHSEN2}. For
earlier discussions of the
possibility of a strong-weak coupling duality in four dimensional
field theory see Refs.\cite{MONTOLIVE,OSBORN}, and in
four dimensional string theory, see Ref.\cite{FONT}.

\section{Symmetry of the Effective Action} \label{s2}

We shall begin this section by carrying out the dimensional
reduction of the $N=1$ supergravity theory coupled to $N=1$
super Maxwell theory from ten dimensions to four dimensions. In
\S\ref{s2.2} we discuss the O(6,22) and SL(2,R) symmetry of the
resulting effective field theory.  We shall see  that O(6,22)
and SL(2,R) symmetries appear on a somewhat different footing;
the former is a symmetry of the effective action, while the
latter is only a symmetry of the equations of motion.  In
\S\ref{s2.3} we shall show that it is possible to give an
alternative formulation of the theory in which SL(2,R) becomes a
symmetry of the action.  Finally in \S\ref{s2.4} we shall show
that the manifestly SL(2,R) invariant formulation of the theory
can be obtained from the dimensional reduction of the dual
formulation of the $N=1$ supergravity theory from ten to four
dimensions. In later sections we shall see that the discrete
SL(2,Z) subgroup of the SL(2,R) group can be identified as the
S-duality group, just as the discrete O(6,22;Z) subgroup of the
O(6,22) group can be identified as the T-duality group\cite{GIVEON}.

\subsection{Dimensional Reduction of the Ten Dimensional Theory
}
\label{s2.1}

We consider heterotic string theory compactified on a six
dimensional torus. The simplest way to derive the low energy
effective action for this theory is to start with the $N=1$
supergravity theory coupled to $N=1$ super Yang-Mills theory in ten
dimensions, and dimensionally reduce the theory from ten to four
dimensions\cite{FERRARA,HASSANSEN,MAHSCH}.
Since at a generic point in the moduli space only the abelian
gauge fields give rise to massless fields in four dimensions, it
is enough to restrict to the U(1)$^{16}$ part of the ten
dimensional gauge group. The ten dimensional action is given by,
\ben \label{new1}
{1\over 32\pi} \int d^{10}z \sqrt{ - G^{(10)}}\,
e^{-\Phi^{(10)}}\Big(R^{(10)} +G^{\ten MN}\p_M\Phi^\ten \p_N\Phi^\ten
\nonumber \\
- {1\over 12} H^{(10)}_{MNP}
H^{(10)MNP} - {1\over 4} F^{(10)I}_{MN} F^{(10)IMN}\Big),
\een
where $G^{(10)}_{MN}$, $B^{(10)}_{MN}$, $A^{(10)I}_M$, and
$\Phi^{(10)}$ are ten dimensional metric, anti-symmetric tensor
field, U(1) gauge fields and the scalar dilaton field
respectively ($0\le M, N \le 9$, $1\le I\le 16$), and,
\ben \label{new2}
F^{(10)I}_{MN} &=& \p_M A^{\ten I}_N - \p_N A^{\ten I}_M \nonumber\\
H^{(10)}_{MNP} &=& (\p_M B^\ten_{NP} -{1\over 2} A_M^{\ten I}
F^{\ten I}_{NP}) + \hbox{cyclic permutations in $M$, $N$, $P$}.
\een
We have ignored the fermion fields in writing down the action
\refb{new1}; we shall discuss them in \S\ref{s2.5}. Also note that we have
included a factor of $(1/32\pi)$ multiplying the action for later
convenience. This factor can be absorbed into $\Phi^\ten$ by shifting it
by $\ln 32\pi$.

For dimensional reduction, it is convenient to introduce the
`four dimensional fields'  $\wh G_{mn}$, $\wh B_{mn}$, $\wh
A^I_m$, $\Phi$, $A_\mu^\al$, $G_{\mu\nu}$ and $B_{\mu\nu}$ ($1\le
m\le 6$, $0\le \mu\le 3$, $1\le a \le 28$) through the
relations\cite{MAHSCH,SEN1,SCHWARZ1}\footnote{The normalization and sign
conventions used here are slightly different from those used in
Ref.\cite{SEN1}. Care has been taken to ensure that we use the same
normalization convention throughout this paper.}
\ben\label{1.2}
&& \wh G_{mn}  = G^\ten_{m+3,n+3}, \quad  \wh B_{mn}  =
B^\ten_{m+3, n+3}, \quad  \wh A^I_m  = A^{\ten I}_{m+3},
\nonumber \\
&& A^{(m)}_\mu  = {1\over 2}\wh G^{mn} G^\ten_{n+3,\mu}, \quad
A^{(I+12)}_\mu = -({1\over 2} A^{\ten I}_\mu - \wh A^I_n
A^{(n)}_\mu), \nonumber \\
&&  A^{(m+6)}_\mu = {1\over 2}
B^\ten_{(m+3)\mu} - \wh B_{mn} A^{(n)}_\mu + {1\over 2}\wh A^I_m
A^{(I+12)}_\mu, \nonumber \\
&& G_{\mu\nu} = G^\ten_{\mu\nu} - G^\ten_{(m+3)\mu} G^\ten_{(n+3)\nu} \wh
G^{mn}, \nonumber \\
&& B_{\mu\nu} = B^\ten_{\mu\nu} - 4\wh B_{mn} A^{(m)}_\mu
A^{(n)}_\nu - 2 (A^{(m)}_\mu A^{(m+6)}_\nu - A^{(m)}_\nu A^{(m+6)}_\mu),
\nonumber \\
&& \Phi = \Phi^\ten - {1\over 2} \ln\det \wh G, \quad \quad
\quad 1\le m, n \le 6, \quad
0\le \mu, \nu \le 3, \quad 1\le I \le 16.
\een
Here $\wh G^{mn}$ denotes the inverse of the matrix $\wh G_{mn}$.
We now combine the scalar fields $\wh G_{mn}$, $\wh B_{mn}$, and
$\wh A_m^I$ into an  $O(6,22)$ matrix valued scalar field $M$.
For this we regard $\wh G_{mn}$, $\wh B_{mn}$ and $\wh A^I_m$ as
$6\times 6$, $6\times 6$, and $6\times 16$ matrices
respectively, and $\wh C_{mn} = {1\over 2} \wh A^I_m
\wh A^I_n$ as a $6\times 6$ matrix, and define $M$ to be the
$28\times 28$ dimensional matrix
\be \label{newa1}
M = \pmatrix{\displaystyle \wh G^{-1} & \wh G^{-1} (\wh B + \wh
C) & \wh G^{-1}\wh A \cr (-\wh B + \wh C) \wh G^{-1} & (\wh G
- \wh B +
\wh C) \wh G^{-1} (\wh G + \wh B + \wh C) & (\wh G -\wh B +\wh
C)\wh G^{-1} \wh A \cr  \wh A^T \wh G^{-1} &  \wh A \wh G^{-1}
(\wh G + \wh B +\wh
C) & I_{16} + \wh A^T \wh G^{-1} \wh A \cr }.
\ee
satisfying
\be \label{1.1}
M L M^T = L, \quad \quad  M^T=M, \quad \quad L =\pmatrix{0 & I_6
& 0  \cr I_6 & 0 & 0 \cr 0 & 0 & -I_{16}},
\ee
where  $I_n$ denotes the $n\times n$ identity matrix.

The effective action that governs the dynamics of the massless
fields in the four dimensional theory is obtained by
substituting the expressions for the ten dimensional fields in
terms of the four dimensional fields in Eq.(\ref{new1}), and taking all
field configurations to be independent of the internal coordinates. The
result is
\ben \label{1.3}
S &=& {1\over 32\pi} \int d^4 x \sqrt{-  G} \, e^{-\Phi} \big[ R_G +
G^{\mu\nu}
\p_\mu \Phi \p_\nu\Phi -{1\over 12} G^{\mu\mu'} G^{\nu\nu'}
G^{\rho\rho'} H_{\mu\nu\rho} H_{\mu'\nu'\rho'} \nonumber \\
&&\quad\quad  - G^{\mu\mu'} G^{\nu\nu'} F^\al_{\mu\nu} (LML)_{ab}
F^\bet_{\mu'\nu'} + {1\over 8} G^{\mu\nu} Tr (\p_\mu M L \p_\nu
M L) \big]
\een
where
\ben \label{1.5}
F^\al_{\mu\nu} &=& \p_\mu A^\al_\nu - \p_\nu A^\al_\mu \nonumber \\
H_{\mu\nu\rho} &=& (\p_\mu B_{\nu\rho} + 2 A^\al_\mu
L_{ a  b } F^\bet_{\nu\rho}) + \hbox{cyclic permutations of
$\mu$, $\nu$, $\rho$},
\een
and $R_G$ is the scalar curvature associated with the four
dimensional metric $G_{\mu\nu}$. In deriving this result we have taken
$\int d^6 y=1$, where $y^m$ ($1\le m\le 6$) denote the coordinates
labeling the six dimensional torus.

\subsection{O(6,22) and SL(2,R) Symmetries of the Effective
Field Theory} \label{s2.2}

This effective action can easily be seen to be invariant under
an $O(6,22)$
transformation\cite{GIVEON}
\be \label{1.3a}
M \to \Omega M \Omega^T, \quad  A_\mu^\al \to \Omega_{ab}
A_\mu^\bet, \quad G_{\mu\nu} \to G_{\mu\nu}, \quad B_{\mu\nu}\to
B_{\mu\nu}, \quad \Phi\to \Phi
\ee
where $\Omega$ is an $O(6,22)$ matrix, satisfying
\be \label{1.3b}
\Omega^T L \Omega = L.
\ee
An $O(6,22;Z)$ subgroup of this is
known to be an exact symmetry of the full string theory and will
be called the T-duality group in this paper. Part of
this symmetry exchanges the Kaluza-Klein modes of the theory,
i.e. the states carrying momenta in the internal directions,
with the string winding modes, $-$ states corresponding to
a string wrapped around one of the compact directions.

The effective four dimensional theory is invariant under another
set of symmetry transformations, which correspond to a symmetry
of the equations of motion, but not of the effective action
given in Eq.(\ref{1.3}). To exhibit this symmetry, we introduce
the canonical matric,
\be \label{1.3aa}
g_{\mu\nu} = e^{-\Phi} G_{\mu\nu},
\ee
and use the convention that all indices are raised or lowered
with respect to this canonical metric. Also, we denote by $D_\mu$ the
standard
covariant derivative constructed from the metric $g_{\mu\nu}$.
The $B_{\mu\nu}$ equations of motion, as derived from the action
\refb{1.3}, are given by
\be \label{1.6}
D_\rho (e^{-2\Phi} H^{\mu\nu\rho}) =0
\ee
which allows us to introduce a scalar field $\Psi$ through the
relation
\be \label{1.7}
H^{\mu\nu\rho} = - (\sqrt{-  g})^{-1} e^{2\Phi}
\epsilon^{\mu\nu\rho\sigma} \p_\sigma \Psi.
\ee
Let us introduce a complex scalar field
\be \label{1.7a}
\lambda = \Psi + i e^{-\Phi}\equiv \lambda_1 + i\lambda_2
\ee
The equations of motion of the fields
$G_{\mu\nu}$, $A_\mu^{( a )}$ and $\Phi$, derived from
the action given in (\ref{1.3}), together with the Bianchi
identity for the field strength $H_{\mu\nu\rho}$, may now be
written as,
\ben \label{1.8}
&& R_{\mu\nu} = {\p_\mu\bar\lambda \p_\nu\lambda +
\p_\nu\bar\lambda \p_\mu\lambda \over 4 (\lambda_2)^2} + 2
\lambda_2 F^\al_{\mu\rho} (LML)_{ab} F^{\bet\rho}_\nu - {1\over
2} \lambda_2 g_{\mu\nu} F^\al_{\rho\sigma} (LML)_{ab}
F^{\bet\rho\sigma}, \nonumber \\
&& D_\mu
(-\lambda_2 (ML)_{ab} F^{\bet\mu\nu} + \lambda_1
\tF^{\al\mu\nu}) =0, \nonumber \\
&& {D^\mu D_\mu\lambda\over
(\lambda_2)^2} + i {D_\mu\lambda D^\mu\lambda \over
(\lambda_2)^3} -i F_{\mu\nu}^\al (LML)_{ab} F^{\bet\mu\nu} +
\tF^\al_{\mu\nu} L_{ab} F^{\bet\mu\nu} =0,
\een
where $R_{\mu\nu}$ is the Ricci tensor calculated with the
metric $g_{\mu\nu}$, and,
\be \label{deftef}
\tF^{\al\mu\nu} = {1\over 2}
(\sqrt{-  g})^{-1}\epsilon^{\mu\nu\rho\sigma} F^\al_{\rho\sigma}.
\ee
Derivation of the equations of motion for the field $M$ is a little bit
more complicated, since $M$ is a constrained matrix. The simplest way to
derive these equations is to introduce a set of independent parameters
$\{\phi^i\}$ that label the symmetric O(6,22) matrix $M$. (We can take
$\phi^i$ to be the set $\{\wh G_{mn}, \wh B_{mn}, \wh A^I_m\}$,
but any other parametrization will also do.) Varying the action with
respect to these parameters $\phi^i$, we get the following set of
equations of motion,
\be \label{emequation}
{1\over 4} Tr\Big( {\delta M\over \delta\phi^i} L D_\mu D^\mu M L\Big) +
\lambda_2 F^\al_{\mu\nu} \Big( L {\delta M\over \delta \phi^i} L\Big)_{ab}
F^{\bet \mu\nu} =0.
\ee
Finally, the Bianchi identities satisfied by the gauge field
strengths $F^{( a )}_{\mu\nu}$ are given by,
\be \label{1.9}
D_\mu \tF^{\al\mu\nu} = 0.
\ee
It is now straightforward to check that the set of equations
(\ref{1.8}), \refb{emequation} and (\ref{1.9}) are invariant under the
following set of
SL(2,R) transformations\cite{DEROO,STW,SEN1,SCHWARZ1}:
\be \label{1.10}
\lambda\to \lambda'= {a\lambda + b\over c\lambda + d}, \quad
F^\al_{\mu\nu}
\to F^{\prime\al}_{\mu\nu} = (c\lambda_1+d)  F^\al_{\mu\nu} +
c\lambda_2 (ML)_{ab}
\tF^\bet_{\mu\nu}, \quad g_{\mu\nu} \to g_{\mu\nu}, \quad M\to M.
\ee
where $a$, $b$, $c$ and $d$ are real numbers satisfying
$ad-bc=1$.
In particular, if we consider the element $a=0$, $b=1$, $c=-1$
and $d=0$, then the transformations take the form:
\be \label{1.11}
\lambda \to -{1\over \lambda}, \quad \quad F^\al_{\mu\nu} \to
-\lambda_1 F^\al_{\mu\nu} - \lambda_2 (ML)_{ab} \tF^\bet_{\mu\nu}.
\ee
For $\lambda_1=0$, this transformation takes electric fields to
magnetic fields and vice versa. It also takes $\lambda_2$ to
$1/\lambda_2$. Since $(\lambda_2)^{-1}=e^{\Phi}$ can be identified
with the coupling constant of the string theory, we see that the duality
transformation takes a strong coupling theory to a weak coupling theory
and vice-versa. We shall refer to the transformations (\ref{1.11}) as the
strong-weak coupling duality transformation, or electric-magnetic duality
transformation.  Note that the full SL(2,R) group of transformations is
generated as a combination of the transformation (\ref{1.11}) and the
trivial duality transformation
\be \label{1.12}
\lambda_1\to \lambda_1+b,
\ee
with all other fields remaining invariant.

It can be easily checked that the set of equations \refb{1.8} and
\refb{emequation} can be derived from the action
\ben \label{2.1}
S &=& {1\over 32\pi} \int d^4 x \sqrt{-  g}\, \big[ R -{1\over
2(\lambda_2)^2} g^{\mu\nu} \p_\mu\lambda \p_\nu\bar\lambda
-\lambda_2 F^\al_{\mu\nu}  (LML)_{ab}   F^{\bet\mu\nu} \nonumber \\
&& + \lambda_1 F^\al_{\mu\nu}  L_{ab}   \tF^{\bet\mu\nu} +{1\over
8} g^{\mu\nu} Tr(\p_\mu M L \p_\nu M L)\big]
\een
This form of the action will be useful to us for later analysis.

We wish to know whether any subgroup of this
SL(2,R) group can be an exact symmetry of string theory,
in the same way that the O(6,22;Z) subgroup of O(6,22) is an exact
symmetry of string theory. However, before we address
this question, we notice that even at the level of effective
action, there is an asymmetry between the
O(6,22) and SL(2,R) symmetry transformtions. The former
is a symmetry of the effective action, whereas the latter is only a
symmetry of the equations of  motion. We shall now show how to
reformulate the theory so that SL(2,R) becomes a symmetry of the
effective action\cite{SCHSEN1,SCHWARZ2}.

\subsection{Manifestly SL(2,R) Invariant Action}\label{s2.3}

We begin by defining the matrices,
\be \label{defcm}
\M = {1\over \lambda_2} \pmatrix{ 1 & \lambda_1 \cr \lambda_1 &
|\lambda|^2\cr}, \quad \quad \LL=\pmatrix{0 & 1 \cr -1 & 0}.
\ee
We also introduce a set of auxiliary gauge
fields\cite{KALORT,SCHSEN1} $A_\mu^{( a,2 )}$ ($1\le a\le 28$),
and define,
\be\label{mmmone}
A_\mu^{(a,1)} = A_\mu^\al,
\ee
\be\label{mmmtwo}
F^{(a,\alpha)}_{\mu\nu} = \p_\mu A_\nu^{(a, \alpha)}- \p_\nu
A_\mu^{(a,\alpha)},
\quad \quad E^{(a,\alpha)}_i = F^{(a,\alpha)}_{0i},\quad \quad
B^{(a,\alpha) i} = \tF^{(a, \alpha)0i} = (\sqrt{-g})^{-1}
\epsilon^{0ijk} \p_j A^{(a, \alpha)}_k,
\ee
for $1\le\alpha \le 2$.
It can be checked that the set of equations (\ref{1.8}), \refb{emequation}
and (\ref{1.9}) are identical to the equations of motion and Bianchi
identities derived from the
action\cite{SCHSEN1}
\ben \label{1.14}
S &=& {1\over 32\pi} \int d^4 x \Big[ \sqrt{-g}\, \big\{ R - {1\over 4}
g^{\mu\nu}
tr(\p_\mu \M \LL \p_\nu \M\LL) +{1\over 8} g^{\mu\nu}
Tr(\p_\mu M L \p_\nu M L) \big\} \nonumber \\
&& \quad \quad -2 \big\{ B^{(a,\alpha)i} \LL_{\alpha\beta} L_{ab}
E_i^{(b,\beta)} +\varepsilon^{ijk} {g^{0k}\over g^{00}} B^{(a, \alpha)i}
\LL_{\alpha\beta} L_{ab} B^{(b, \beta)j} \nonumber \\
&& \quad \quad -{g_{ij}\over \sqrt{-g} g^{00}} B^{(a,\alpha)i} (\LL^T \M
\LL)_{ \alpha\beta} (LML)_{ab} B^{(b, \beta)j}\big\}\Big].
\een
where $tr$ and $Tr$ denote traces over the indices $\alpha,
\beta$ and $a,b$ respectively.
The simplest way to check that this action gives rise to the
same set of equations as (\ref{1.8}), \refb{emequation} and (\ref{1.9}) is
to note that the $A_i^{( a, 2 )}$ equations of motion give
\be \label{1.14a}
\varepsilon^{ijk} \p_j \Big[ L_{ab} E_k^{(b,1)} +\varepsilon^{klm}
{g^{0m}\over g^{00}} L_{ab} B^{(b,1)l}
+{g_{kl}\over \sqrt{-g} g^{00}} (\M\LL)_{1\beta}
(LML)_{ab} B^{(b,\beta)l}\Big] =0
\ee
where $\varepsilon^{ijk}=\epsilon^{0ijk}$ is the three
dimensional totally anti-symmetric tensor density.  Since these
equations do not involve any time derivative of the fields
$A_i^{( a,2 )}$, we can treat $A_i^{( a,2 )}$ as
auxiliary fields, and eliminate them from the action
(\ref{1.14}) by using their equations of motion.  The resulting
action is identical to the action \refb{2.1}.

The action \refb{1.14} is invariant under manifest SL(2,R)
transformation
\be \label{1.15}
\M\to \omega \M \omega^T, \quad \quad A^{(a,\alpha)}_\mu\to
\omega_{\alpha\beta}  A^{(a,\beta)}_\mu,
\ee
and O(6,22) transformations
\be \label{1.16}
M\to \Omega M \Omega^T, \quad \quad A^{(a,\alpha)}_\mu \to
\Omega_{ab} A^{(b, \alpha)}_\mu,
\ee
where
\be\label{defomega}
\omega = \pmatrix{ d & c \cr b & a\cr}, \quad \quad ad-bc=1,
\ee
is an SL(2,R) matrix, satisfying,
\be\label{omegareln}
\omega^T \LL \omega = \LL.
\ee
The transformation laws of $\lambda$, induced by
Eq.\refb{1.15},
can be seen to be identical to those given in Eq.\refb{1.10}. Also, after
we eliminate the fields $A_i^\alt$ by
their equations of motion, the O(6,22) and SL(2,R)
transformation laws of the rest of the fields coincide with
those given in Eqs.\refb{1.3a} and \refb{1.10}.  The
loss of manifest SL(2,R) invariance of the action after
integrating out the gauge field components $A_i^\alt$ can be
traced to the fact that the set of fields $A_i^\alt$ is not an
SL(2,R) invariant set, since they transform to linear
combinations of $A_i^{(a,1)}$ and $A_i^\alt$ under SL(2,R)
transformations. In contrast, this set is invariant under O(6,22)
transformation, since the fields in this set transform to linear
combinations of the fields in the same set under O(6,22)
transformations.

The action \refb{1.14} is also invariant under the gauge
transformations
\be \label{1.17}
\delta A^{(a,\alpha)}_\mu = \p_\mu \Lambda^{(a, \alpha)}, \quad \quad
\delta A^{(a, \alpha)}_0 = \Psi^{(a, \alpha)},
\ee
where $\Lambda^{(a, \alpha)}$ and $\Psi^{(a, \alpha)}$ are the
gauge transformation parameters.  Note that the action does not
depend on $A_0^{(a, \alpha)}$.  Finally, although \refb{1.14} is
not manifestly general coordinate invariant, it is invariant
under a hidden `general coordinate transformation'
\ben \label{1.18}
\delta A_i^{(a,\alpha)} &=& \xi^j \p_j A_i^{(a, \alpha)} + (\p_i \xi^j)
A^{(a,\alpha)}_j \nonumber \\
&& -\xi^0 \Big\{ {g_{ij}\over \sqrt{-g} g^{00}} (\M\LL)_{\alpha\beta}
(ML)_{ab} B^{(b, \beta)j} + {g^{0k}\over g^{00}} \varepsilon^{ijk}
B^{(a,\alpha)j}\Big\}, \nonumber \\
\delta M = \xi^\mu \p_\mu M, && \quad \delta \M=\xi^\mu \p_\mu \M,
\quad \quad \delta g_{\mu\nu} = \xi^\rho \p_\rho g_{\mu\nu} + g_{\rho\nu}
\p_\mu \xi^\rho + g_{\mu\rho} \p_\nu \xi^\rho.
\een
This transformation does not look like the usual general
coordinate transformation. However, if we use the equations of
motion of $A_i^\alt$ given in \refb{1.14a}, the transformation
laws of all other fields reduce to the usual general coordinate
transformation laws\cite{SCHSEN1}.

Thus we see that the low energy effective theory of the four
dimensional heterotic string can be described by a manifestly
SL(2,R)$\times$O(6,22) invariant action. This action is not
manifestly general coordinate invariant, but has a hidden
general coordinate invariance. One can now ask if it is possible
to find another action describing the same theory, which is
manifestly SL(2,R) and general coordinate invariant. It turns
out that this is possible for a restricted class of
configurations where we set all fields originating from the
ten dimensional gauge fields $A^{(10)I}_M$ to zero\cite{SCHSEN1}. In terms
of four dimensional fields this would correspond to replacing the 28
component gauge field $A_\mu^\al$ by a 12 component gauge field $\wc
A^\bet_\mu$ ($1\le b \le 12$), and $M$ by  a 12$\times$12 matrix $\wc M$,
satisfying,
\be \label{1.19}
\wc M^T=\wc M, \quad \quad \wc M \wc L \wc M^T = \wc L, \quad \quad  \wc L
=\pmatrix{0 & I_6 \cr I_6 & 0}.
\ee
The action \refb{2.1} is now replaced by,
\ben \label{1.20}
S &=& {1\over 32\pi} \int d^4 x \sqrt{-  g}\, \big[ R -{1\over
2(\lambda_2)^2} g^{\mu\nu} \p_\mu\lambda \p_\nu\bar\lambda
-\lambda_2 \wc F^\al_{\mu\nu}  (\wc L\wc M\wc L)_{ab}   \wc F^{\bet\mu\nu}
\nonumber \\
&& + \lambda_1 \wc F^\al_{\mu\nu}  \wc L_{ab}   \wc {\tF}^{\bet\mu\nu}
+{1\over
8} g^{\mu\nu} Tr(\p_\mu \wc M \wc L \p_\nu \wc M \wc L)\big]
\een
where
\be \label{1.5new}
\wc F^\al_{\mu\nu} = \p_\mu \wc A^\al_\nu - \p_\nu \wc
A^\al_\mu.
\ee
The indices $a,b$ run from 1 to 12.  This action has manifest
O(6,6) symmetry. As in the previous case, the equations of
motion are invariant under SL(2,R) transformation, but the
effective action is not SL(2,R) invariant. As before, this theory may be
shown to be equivalent to a manifestly SL(2,R) and O(6,6) invariant, but
not manifestly general coordinate invariant, action
\ben \label{1.20a}
S &=& {1\over 32\pi} \int d^4 x \Big[\sqrt{-g}\, \big\{ R - {1\over 4}
g^{\mu\nu}
tr(\p_\mu \M \LL \p_\nu \M\LL) +{1\over 8} g^{\mu\nu}
Tr(\p_\mu \wc M \wc L \p_\nu \wc M \wc L) \big\} \nonumber \\
&& \quad \quad - 2 \big\{ \wc B^{(a,\alpha)i} \LL_{\alpha\beta} \wc L_{ab}
\wc E_i^{(b,\beta)} +\varepsilon^{ijk} {g^{0k}\over g^{00}}
\wc B^{(a, \alpha)i}
\LL_{\alpha\beta} \wc L_{ab} \wc B^{(b, \beta)j} \nonumber \\
&& \quad \quad -{g_{ij}\over \sqrt{-g} g^{00}} \wc B^{(a,\alpha)i} (\LL^T
\M \LL)_{
\alpha\beta} (\wc L\wc M\wc L)_{ab} \wc B^{(b, \beta)j}\big\}\Big].
\een
The SL(2,R) and O(6,6) transformations act on the various fields
as
\be \label{1.20b}
\M\to \omega \M \omega^T, \quad\quad \wc M\to \wc\Omega \wc M \wc\Omega^T,
\quad \quad \wc A^{(a, \alpha)}_\mu \to \omega_{\alpha\beta}
\wc\Omega_{ab} \wc A^{(b, \beta)},
\ee
where $\wc\Omega$ is an O(6,6) matrix satisfying $\wc\Omega \wc L
\wc\Omega^T = \wc L$.  If we eliminate the O(6,6) invariant set
of fields $\wc A_i^{(b,2)}$ for $1\le b\le 12$ by their
equations of motion, we recover the original action
\refb{1.20}. Instead of doing that, we can also eliminate  the
SL(2,R) invariant set of fields $\wc A_i^{(m+6,\beta)}$ for
$1\le \beta\le 2$, and
$1\le  m \le 6$ by their equations of motion, since these
equations do not contain any time derivative of these fields.
The resulting action is\cite{SCHSEN1}
\ben \label{1.21}
{1\over 32\pi} \int d^4 x \sqrt{-g} \Big[ R - {1\over 4} g^{\mu\nu}
tr(\p_\mu \M
\LL \p_\nu\M \LL) +{1\over 8} g^{\mu\nu} Tr(\p_\mu \wc M \wc L
\p_\nu \wc M\wc L) && \nonumber \\
- \wc F^{(m,\alpha)}_{\mu\nu} \wh G_{mn} (\LL^T\M\LL)_{\alpha
\beta} \wc F^{(n,\beta)}_{\rho\sigma} g^{\mu\rho} g^{\nu\sigma}
-\wc F^{(m, \alpha)}_{\mu\nu} \wh B_{mn} \LL_{\alpha\beta}
\wc{\tF}^{(n,\beta)}_{\rho\sigma} g^{\mu\rho} g^{\nu\sigma}\Big], &&
\nonumber \\
1\le m,n\le 6, &&
\een
which is manifestly  general coordinate and SL(2,R) invariant,
but is not O(6,6) invariant. The equations of motion,
however, are invariant under the O(6,6)
transformations.\footnote{ Note that this procedure cannot be
carried out for the action \refb{1.14}, since in that case we
cannot find an SL(2,R) invariant set of fields whose equations
of motion do not contain time derivative of the fields being
eliminated.}

Thus we see that at the level of the effective action, we have
been able to put O(6,6)$\big($O(6,22)$\big)$ transformations and the
SL(2,R) transformations on an equal footing. First, there is a
formulation of the theory in which O(6,22) is a manifest
symmetry of the action whereas SL(2,R) is only a symmetry of the
effective action. Second, there is a different formulation of
the theory where the action is manifestly O(6,22) and SL(2,R)
invariant, but not manifestly general coordinate invariant.
Finally, in the special case when we ignore the ten dimensional
gauge fields, there is a third formulation of the theory where
the action is manifestly SL(2,R) and general coordinate
invariant, but O(6,6) is only a symmetry of the equations of
motion.

Despite these three alternate formulations of the action, one of
them, namely \refb{1.3}, appears to be more fundamental, since
this is the action that comes from the dimensional reduction of
the $N=1$ supergravity action in ten dimensions. We shall now
show that if we start with the dual formulation of the $N=1$
supergravity theory in ten dimensions, then we recover a
manifestly SL(2,R) invariant form of the action after dimensional
reduction\cite{SCHSEN1,BINET}.

\subsection{Manifestly SL(2,R) Invariant Effective Action from
Dimensional Reduction of the Dual $N=1$ Supergravity Theory in
Ten Dimensions}\label{s2.4}

The dual formulation of the $N=1$ supergravity theory in ten
dimensions is based on the metric $\wt G^\ten_{MN}$, a six-form field
$\wt B^\ten_{M_1\ldots M_6}$, and the dilaton field
$\wt\Phi^\ten$. (We are ignoring the ten dimensional gauge
fields and the fermionic fields in the analysis of this
section.) The action is given by\cite{DUFFLUONE},
\ben \label {1.22}
S &=& {1\over 32\pi} \int d^{10} z \sqrt{- \wt G^\ten}\,
e^{\wt\Phi^\ten/3}
\Big( \wt R^\ten \nonumber \\
&& \quad \quad - {1\over 2 \times 7!} \wt G^{\ten M_1 N_1} \cdots \wt
G^{\ten M_7 N_7}
\wt H^\ten_{M_1\ldots M_7} \wt H^\ten_{N_1 \ldots N_7}\Big),
\een
where
\be \label{1.23}
\wt H^\ten_{M_1 \ldots M_7} = \p_{[M_1} \wt B^\ten_{M_2 \ldots M_7]}.
\ee
The equations of motion and the Bianchi identities derived from this
action can be shown to be identical to those derived from the action
\refb{new1} provided we make the identifications
\ben \label{1.23a}
&& \wt\Phi^\ten =\Phi^\ten, \quad \quad \wt G^\ten_{MN} = e^{-\Phi^\ten/3}
G^\ten_{MN}, \nonumber \\
&& \sqrt{- \wt G^\ten}\, e^{\wt \Phi^\ten/3} \wt G^{\ten M_1 N_1} \cdots
\wt G^{\ten M_7 N_7} \wt H^\ten_{N_1 \ldots N_7}
= - {1\over 3!} \epsilon^{M_1 \ldots M_{10}} H_{M_8 M_9 M_{10}}.
\een
Note that the Bianchi identity for the field strength $H^\ten_{MNP}$
in ten dimensions
\be \label{1.24}
\epsilon^{M_1\ldots M_{10}} \p_{M_7} H^\ten_{M_8 M_9 M_{10}} =0,
\ee
now corresponds to the equation of motion for the six form field
$\wt B^\ten_{M_1\ldots M_6}$. Similarly, the Bianchi identity for the
field strength $\wt H_{M_1\ldots M_7}$ in ten dimensions
\be \label{1.25}
\epsilon^{M_1\ldots M_{10}} \p_{M_3} \wt H^\ten_{M_4 \ldots M_{10}} =0,
\ee
corresponds to the equation of motion of the anti-symmetric tensor
field $B^\ten_{MN}$.

In order to carry out the dimensional reduction of this theory
from ten to four dimensions, it is convenient to introduce the
`four dimensional fields' $\lambda$, $\C_\mu^m$, $\D_\mu^m$,
$\wh G_{mn}$, $\B_{\mu\nu}^{mn}$, $\E_{\mu\nu\rho}^{mnp}$ and
$g_{\mu\nu}$ through the relations\cite{SCHSEN1}:
\ben \label{1.26}
\wh G_{mn} &=& e^{\wt \Phi^\ten/3} \wt G^\ten_{m+3,n+3}, \quad \quad
\lambda_1 = {1\over 6!}
\wt B^\ten_{m_1+3, \ldots m_6+3} \epsilon^{m_1\ldots m_6},
\quad \quad \lambda_2 = \sqrt{\det\wh G}\,  e^{-\wt \Phi^\ten},
\nonumber \\
\C^m_\mu &=& e^{\wt \Phi^\ten/3} \wh G^{mn} \wt G^\ten_{(n+3)\mu},
\quad \quad
\D^{m_1}_\mu = {1\over 5!} \epsilon^{m_1\ldots m_6} \wt B^\ten_{
\mu (m_2+3)\ldots (m_6+3)} - \lambda_1 \C^{m_1}_\mu
\nonumber \\
\B^{m_1 m_2}_{\mu\nu} &=& {1\over 4!} \epsilon^{m_1\ldots m_6}
\wt B^\ten_{\mu\nu (m_3+3) \ldots (m_6+3)} \nonumber \\
&&  - [(\lambda_1 \C^{m_1}_\mu
\C^{m_2}_\nu +{1\over 2} \D^{m_1}_\mu \C^{m_2}_\nu -{1\over 2}
\D^{m_1}_\nu \C^{m_2}_\mu) - (m_1\leftrightarrow m_2)]\nonumber \\
\E^{m_1 m_2 m_3}_{\mu\nu\rho} &=& {1\over 3!} \epsilon^{m_1 \ldots m_6}
\wt B^\ten_{\mu\nu\rho (m_4+3) \ldots (m_6+3)}, \nonumber \\
g_{\mu\nu} &=& (\lambda_2)^{2/3} (\det\wh G)^{{1\over 6}}
(\wt G^\ten_{\mu\nu} - \wt G^\ten_{(m+3)(n+3)} \C^m_\mu \C^n_\nu),
\een
and the corresponding field strengths,
\ben \label{1.27}
F^{(\C)m}_{\mu\nu} &=& \p_\mu \C^m_\nu - \p_\nu \C^m_\mu, \quad \quad
F^{(\D)m}_{\mu\nu} = \p_\mu \D^m_\nu - \p_\nu \D^m_\mu \nonumber \\
K^{mn}_{\mu\nu\rho} &=&  \Big(\big[ \p_\mu \B^{mn}_{\nu\rho}
-{1\over 2} \big\{ (\C^n_\rho F^{(\D)m}_{\mu\nu} + \D^n_\rho F^{(\C)m}_{
\mu\nu}) - (m\leftrightarrow n)\big\}\big] \nonumber \\
&&  + \hbox{ cyclic permutations of $\mu$, $\nu$, $\rho$}\Big)
\nonumber \\
\K^{mnp}_{\mu\nu\rho\sigma} &=& \big[ \p_\mu \E^{mnp}_{\nu\rho\sigma}
+(-1)^P \cdot\hbox{ cyclic permutations of $\mu$, $\nu$, $\rho$, $\sigma$}]
\nonumber \\
&&  - \big[ (\C^p_\sigma K^{mn}_{\mu\nu\rho} + \hbox{ cyclic
permutations of $m$, $n$, $p$})\nonumber \\
&&  +(-1)^P \cdot \hbox{ cyclic permutations of $\mu$, $\nu$,
$\rho$, $\sigma$}\big]\nonumber \\ && -\big[ \big\{ \C^p_\sigma
\C^n_\rho (F^{(\D)m}_{\mu\nu} + \lambda_1 F^{(\C)m}_{\mu\nu}) +
(-1)^P \cdot \hbox{ all permutations of $m$, $n$, $p$}\big\}
\nonumber \\
&&  +(-1)^P \cdot \hbox{ inequivalent permutations
of $\mu$, $\nu$, $\rho$, $\sigma$}\big] \nonumber \\
&&  -\big[ (
\C^p_\sigma \C^n_\rho \C^m_\nu \p_\mu \lambda_1 + (-1)^P \cdot
\hbox{all permutations of $m$, $n$, $p$}) \nonumber \\
&& +(-1)^P \cdot \hbox{ cyclic permutations of $\mu$, $\nu$,
$\rho$, $\sigma$}\big].
\een
Using the relationship between the fields in the two
formulations of the ten dimensional $N=1$ supergravity theory
given in Eq.\refb{1.23a}, and the definition of the fields
$\lambda_1$, $\lambda_2$, $\wh G_{mn}$ and $g_{\mu\nu}$ in the
two formulations, one can easily verify that the two sets of
definitions lead to identical $\lambda$, $\wh G_{mn}$ and
$g_{\mu\nu}$.

The action \refb{1.22}, expressed in terms of these `four
dimensional fields', is given by,
\ben\label{1.28}
S &=& {1\over 32\pi} \int d^4 x \sqrt{-g} \Big[ R - {1\over 2
(\lambda_2)^2}
g^{\mu\nu} \p_\mu \bar\lambda \p_\nu\lambda +{1\over 4} g^{\mu\nu}
\hbox{Tr}(\p_\mu\wh G \p_\nu \wh G^{-1}) \nonumber \\
&& -{1\over 4} \wh G_{mn} g^{\mu\rho} g^{\nu\sigma}
\pmatrix{F^{(\C)m}_{\mu\nu
} & -F^{(\D)m}_{\mu\nu}} \LL^T\M \LL
\pmatrix{F^{(\C)n}_{\rho\sigma} \cr
- F^{(\D)n}_{\rho\sigma}}\nonumber \\
&& - {1\over 2\times 2! \times 3!} \wh G_{m_1 n_1} \wh G_{m_2 n_2}
g^{\mu_1\nu_1}\cdots g^{\mu_3\nu_3} K^{m_1m_2}_{\mu_1\mu_2\mu_3}
K^{n_1n_2}_{\nu_1\nu_2\nu_3}\nonumber \\
&& - {\lambda_2 \over 2 \times 3! \times 4!} \wh G_{m_1n_1}\cdots \wh
G_{m_3n_3}
g^{\mu_1\nu_1}\cdots g^{\mu_4\nu_4} \K^{m_1\ldots m_3}_{\mu_1\ldots \mu_4}
\K^{n_1\ldots n_3}_{\nu_1\ldots \nu_4}\Big],
\een
where $\M$ has been defined in Eq.\refb{defcm}, and Tr denotes trace over
the indices $m,n$ ($1\le m,n\le 6$).
The equation of motion for $\E^{m_1m_2m_3}_{\mu_1\mu_2\mu_3}$ gives
\be \label{1.29}
\p_{\nu_1}\big[\lambda_2\sqrt{-g}\, \wh G_{m_1n_1} \ldots \wh G_{m_3n_3}
g^{\mu_1\nu_1}\ldots g^{\mu_4\nu_4} \K^{n_1\ldots n_3}_{\nu_1\ldots \nu_4}
\big]=0.
\ee
Since $\K^{n_1\ldots n_3}_{\nu_1\ldots \nu_4}$ is antisymmetric in
$\nu_1, \ldots \nu_4$, we may write
\be \label{1.30}
\lambda_2\sqrt{-g} \, \wh G_{m_1n_1} \ldots \wh G_{m_3n_3}
g^{\mu_1\nu_1}\ldots g^{\mu_4\nu_4} \K^{n_1\ldots n_3}_{\nu_1\ldots \nu_4}
=\epsilon^{\mu_1\ldots \mu_4} H_{m_1m_2m_3}
\ee
for some $H_{mnp}$. The
equation \refb{1.29} then takes the form:
\be \label{1.31}
\p_\nu H_{m_1m_2m_3} = 0,
\ee
showing that $H_{mnp}$ is a constant. Comparison with the original
formulation of the theory shows that $H_{mnp}$ are proportional to the
internal components of the three form field strength $H^{(10)}_{MNP}$.
During the dimensional reduction of the original ten dimensional $N=1$
supergravity theory, we had set these constants to zero. Hence, if we want
to recover the same theory, we must set them to zero here too. This gives
\be \label{1.32}
\K^{m_1\ldots m_3}_{\mu_1\ldots \mu_4} =0.
\ee
The action \refb{1.28} now reduces to
\ben \label{1.33}
S &=& {1\over 32\pi} \int d^4 x \sqrt{-g} \Big[ R - {1\over 4}
g^{\mu\nu} tr(\p_\mu \M \LL \p_\nu\M \LL)
+{1\over 4} g^{\mu\nu}
\hbox{Tr}(\p_\mu\wh G \p_\nu \wh G^{-1}) \nonumber \\
&& -{1\over 4} \wh G_{mn} g^{\mu\rho} g^{\nu\sigma}
\pmatrix{F^{(\C)m}_{\mu\nu
} & -F^{(\D)m}_{\mu\nu}} \LL^T\M \LL
\pmatrix{F^{(\C)n}_{\rho\sigma} \cr
- F^{(\D)n}_{\rho\sigma}}\nonumber \\
&& - {1\over 2\times 2! \times 3!} \wh G_{m_1 n_1} \wh G_{m_2 n_2}
g^{\mu_1\nu_1}\cdots g^{\mu_3\nu_3} K^{m_1m_2}_{\mu_1\mu_2\mu_3}
K^{n_1n_2}_{\nu_1\nu_2\nu_3}\Big],
\een
and has manifest SL(2,R) invariance
\be \label{1.34}
\M\to \omega \M \omega^T, \quad \quad \pmatrix{\C^m_\mu \cr -\D^m_\mu}
\to \omega \pmatrix{\C^m_\mu \cr -\D^m_\mu},
\ee
with all other fields remaining invariant under the SL(2,R)
transformation.
Although this action is not identical to the manifestly SL(2,R)
invariant action \refb{1.21}, the equations of motion derived
from these two actions can be seen to be identical, provided we
make the identification
\ben \label{1.35}
&& \sqrt{-g}\, \wh G_{m_1n_1} \wh G_{m_2n_2} g^{\mu_1\nu_1} \cdots
g^{\mu_3\nu_3} K^{m_1m_2}_{\mu_1\mu_2\mu_3} =
- \epsilon^{\nu_1\nu_2\nu_3\sigma} \p_\sigma \wh B_{n_1 n_2},
\nonumber \\
&& \C^m_\mu = 2 \wc A_\mu^{(m,1)}, \quad \quad \D^m_\mu = - 2
\wc A_\mu^{(m,2)}.
\een
Under this identification, the equations of motion of the scalar
field $\wh B_{mn}$ becomes identical to the Bianchi identity of the
field strength $K_{\mu\nu\rho}^{mn}$, and the bianchi identity
of $\p_\mu B_{mn}$ becomes identical to the equations of motion
of the field $\B^{mn}_{\mu\nu}$.

This shows that the SL(2,R) symmetry arises naturally in the
four dimensional theory obtained from the dimensional reduction
of the dual formulation of the $N=1$ supergravity theory in ten
dimensions, just as the O(6,6) or O(6,22) symmetry arises
naturally in the dimensional reduction of the usual $N=1$
supergravity theory from ten to four dimensions. Yet, the O(6,22)
symmetry is more fundamental from the point of view
of string theory, since the fields $G^\ten_{MN}$, $B^\ten_{MN}$,
which arise in the {\it usual} formulation of the N=1 supergravity theory,
couple naturally to the string. On the other hand, it is
known\cite{DUFFLUTWO} that the
fields $\wt G^\ten_{MN}$ and $\wt B^\ten_{M_1\ldots M_6}$ couple
naturally to the five-brane, which has been conjectured to
be equivalent to the theory of
strings\cite{DUFF,STROM,DUFFLUONE,DUFFLUTWO}. Hence one would
expect that the SL(2,R)
symmetry will play a more fundamental role in the theory of five-branes.
In \S\ref{s4} we shall show that there is a natural interpretation of the
SL(2,Z) subgroup of SL(2,R) as the group of target space duality
transformations in the five-brane theory.

\subsection{Inclusion of the Fermions}\label{s2.5}

So far we have concentrated on the bosonic part of the action. However, in
order to establish the SL(2,Z) invariance of the full string theory, it is
necessary to show that the low energy effective field theory is SL(2,Z)
invariant even after inclusion of the massless fermionic fields in the
theory. For this we need to carry out the
dimensional reduction of the full action of the ten dimensional N=1
supergravity theory,
and show the SL(2,Z) invariance of the equations of motion derived from
this action. We shall not do this here. However, we shall give an indirect
argument showing that the equations of motion do remain SL(2,R) invariant
after inclusion of the fermionic fields. This is done by comparing the
dimensionally reduced theory to
the N=4 Poincare supergravity theory coupled to abelian gauge field
multiplets\cite{DEROO}. It can be seen that the bosonic part of the two
theories are identical
if we make the identification\cite{SENSLTZ,FERRARA}
\be\label{deroo1}
M = U OO^T U^{-1}, \quad \quad {i\over \lambda} = {\phi_1 - \phi_2\over
\phi_1 + \phi_2},
\ee
and a redefinition of the gauge fields $F^\al \to U_{ab} F^\bet$. Here
$U$ is the matrix that diagonalizes $L$ to $(I_6, -I_{22})$, and $O$,
$\phi_1$, $\phi_2$ are fields defined in Ref.\cite{DEROO}. Since the
bosonic parts of the two theories are identical, and furthermore, both the
theories have local $N=4$ supersymmetry, we have a strong evidence that
the two theories are indeed the same. We shall proceed with the assumption
that this is the case.

It was shown in Ref.\cite{DEROO} that the gauge field equations of
motion in the Poincare supergravity theory are invariant under SL(2,R)
transformation, even after including the fermionic fields. There is also a
general argument due to Gaillard and Zumino\cite{GAZU}, that if the
gauge field equations in a theory have an SL(2,R) symmetry, then all other
equations of motion also have this symmetry. {}From this we can conclude
that the full set of equations of motion in the N=4 Poincare supergravity
theory, and hence also in the dimensionally reduced low energy heterotic
string theory, are invariant under SL(2,R) transformation.

\section{Symmetry of the Charge Spectrum} \label{s3}

In this section we shall analyze the possibility that part
of the SL(2,R) symmetry can be realised as an exact symmetry of
the theory. Thus the first question that we need to answer is,
which part of SL(2,R) has a chance of being a symmetry of the
full quantum theory. We shall see in \S\ref{s3.1} that the SL(2,R)
symmetry group is necessarily broken down to SL(2,Z) due to the instanton
corrections. Hence the question is whether this SL(2,Z) group of
transformations can be a symmetry group of the full quantum string theory.
As pointed out in the introduction, we shall refer to this group of
SL(2,Z) transformations as the S-duality transformation, and the target
space duality group O(6,22;Z) as the T-duality transformation.

We have already stated that since the S-duality transformation acts
non-trivially on the coupling constant, it is not a symmetry of the theory
order by order in the string perturbation theory, but could only be a
symmetry of the full string theory.  Thus, in order to test this symmetry
we must look for quantities which can be calculated in the full string
theory and see if those quantities are invariant under this symmetry
transformation. We have pointed out in the introduction that there are
four sets of such quantities. Of these, the low energy effective action
has already been shown to possess the SL(2,Z) invariance. In \S\ref{s3.2},
we shall study the SL(2,Z) transformation properties of the allowed
spectrum of electric and magnetic charges in the theory and show that this
spectrum is invariant under the S-duality transformation.

\subsection{Breaking of SL(2,R) to SL(2,Z)}\label{s3.1}

In \S\ref{s2} we wrote down the effective action of the
four dimensional theory in various different forms.
{}From Eq.\refb{2.1} we see that the field $\lambda_1$ couples to
the topological density $F^\al_{\mu\nu}   L_{ab}
\tF^{\bet\mu\nu}$, and hence the part of the SL(2,R) group that
corresponds to a translation symmetry of $\lambda_1$ must be
broken down to a discrete group of translations by
instanton effects. Actually we have to be somewhat
careful, since so far we have introduced only abelian gauge
fields in the theory which do not have any instantons. However,
we should keep in mind that the full string theory contains
non-abelian gauge fields as well. The non-abelian group is
spontaneously broken at a generic point of the moduli space, but
nevertheless gives rise to instanton corrections to the theory.
(At special points in the moduli space, {\it e.g.}, where some of
the $\wh A^I_m$ vanish, part of the non-abelian symmetry group
is restored.) Thus to find how the instanton effects modify the
translation symmetry of $\lambda_1$, we must first study the
embedding of the abelian gauge group in the non-abelian group,
and then compute the (quantized) topological charge that couples
to the zero mode of the field $\lambda_1$.

To take a concrete case, note that the gauge field
$A_\mu^{(28)}$ can be regarded as the gauge field associated
with one of the three generators of an SU(2) group, such that the
unbroken phase of this SU(2) group is restored when $\wh
A^{16}_m$ vanishes for all $m$. Let $\A_\mu^i$ ($1\le i\le 3$) denote
these SU(2) gauge fields. Using the scaling freedom $\lambda\to
c\lambda$, $F^\al_{\mu\nu}\to {1\over\sqrt c} F^\al_{\mu\nu}$, under
which the action remains invariant, we can always ensure that
the field $A_\mu^{(28)}$ is equal to
$\sqrt{2} \A_\mu^3$. Let us assume that this has
been done. In that case, the $-{1\over 32\pi} \int d^4x
\sqrt{-g}\, \lambda_1 \,
\tF^{(28)}_{\mu\nu}   F^{(28)\mu\nu}$ term in the action
can be regarded as a part of the term
\be\label{instcoup}
-{1\over 16\pi} \int d^4 x \sqrt{-  g}\, \lambda_1\, \sum_{i=1}^3
\wt\F^i_{\mu\nu} \F^{i\mu\nu}
\ee
where $\F^i_{\mu\nu}$ are the components of the SU(2) field strength,
\be\label{1nst2}
\F^i_{\mu\nu} = \p_\mu \A^i_\nu -\p_\nu \A^i_\mu +\varepsilon^{ijk}
\A^j_\mu \A^k_\nu.
\ee
Now, it is well known that for a single SU(2) instanton,
\be \label{inst3}
{1\over 16\pi} \int d^4 x \sqrt{-  g}  \sum_{i=1}^3
\tF^i_{\mu\nu} F^{i\mu\nu} = 2\pi.
\ee
As a result, $e^{iS}$ remains invariant under $\lambda_1\to
\lambda_1+\hbox{integer}$.
Thus the presence of this instanton in the theory breaks the
continuous translation symmetry of $\lambda$ to $\lambda\to
\lambda+1$\cite{STW,SEN2}.

One can verify that the $\lambda\to \lambda+1$ symmetry survives the
effect of all other non-abelian instantons in the theory.  Furthermore,
one can show that the subgroup of SL(2,R), generated by the transformation
$\lambda\to \lambda+1$, and the strong-weak coupling duality
transformation $\lambda \to -1/\lambda$, is SL(2,Z). This corresponds to
the subgroup of the SL(2,R) group of transformations generated by matrices
of the form $\pmatrix{a & b\cr c & d}$ with $a$, $b$, $c$, $d$ integers
and satisfying $ad-bc=1$. The effect of these transformations on the
various fields in the theory is the same as that given in Eq.\refb{1.10}.

In the rest of this section we shall find out whether SL(2,Z) can be
an exact symmetry of the charge spectrum of the full string theory.

\subsection{SL(2,Z) Invariance of the Electric and Magnetic
Charge Spectrum}\label{s3.2}

So far in our analysis we have only analyzed the effective
action involving the neutral massless fields in the theory. The full
string theory, of course, also contains charged fields (of which the
non-abelian gauge fields discussed in the previous subsection
are examples). Although at a generic point in the moduli space
of compactification these fields are all massive, and hence
decouple from the low energy effective field theory, we must
show that the spectrum and the interaction of these charged
fields remain invariant under the SL(2,Z) transformation, in
order to establish the SL(2,Z) invariance of the full string
theory.

We start by analyzing the charge spectrum of the states in
string theory\cite{SEN2,KALORT}. In the presence of
charged fields, the fields $A_\mu^\al$ acquire new coupling in
the action of the form
\be \label{amucoup}
- {1\over 2} \int d^4 x \sqrt{-  g} A_\mu^\al(x) J^{\al \mu}(x)
\ee
where $J_\mu^\al$ is the electric current associated with the
charged fields. (The normalization factor of $-{1\over 2}$ is purely a
matter of convention.) Let $e^\al$ be the conserved charge
associated with this current,
\be \label{defea}
e^\al = \int \sqrt{-  g} J^{\al 0} d^3 x.
\ee
We also define the quantity $Q_{el}^\al$ through the relation
\be \label{defqea}
F^\al_{0r} \simeq {Q_{el}^\al\over r^2} \quad \quad \hbox{for large
$r$.}
\ee
Using the equations of motion derived from the sum of the
actions \refb{2.1} and \refb{amucoup}, we see that
\be \label{relqe}
Q_{el}^\al = {1\over \lambda_2^\zero} M^\zero_{ab} e^\bet,
\ee
where the superscript $\zero$ denotes the asymptotic values of
the various fields.

{}From the analysis of Narain\cite{NARAIN}, we know that the
allowed set of electric charge vectors $\{e^\al\}$ are
proportional to vectors $\{\alpha^a\}$ belonging to an even, self
dual, Lorenzian lattice $\Lambda$ with metric $L$ defined in
Eq.\refb{1.1}.\footnote{We can, for definiteness, take $\Lambda$ to be
the direct product of the root lattice of E$_8\times$E$_8$ and the 12
dimensional lattice of integers.}  The constant of proportionality is fixed
as follows. On the one hand, from the analysis of Ref.\cite{NARAIN} we
know that the states associated with the quanta of SU(2) gauge fields
$\A^{\pm}_\mu$ have electric charge vectors of (length)$^2=-2$. On the
other hand, knowing that the Lagrangian of the SU(2) Yang-Mills theory is
proportional to $\F^i_{\mu\nu} \F^{i\mu\nu}$, and using the relation
$A_\mu^{(28)}=\sqrt{2} \A_\mu^3$, and the definition of $e^\al$ given in
Eqs.\refb{amucoup} and \refb{defea}, we can calculate $e^\al$ for the
quanta of states created by the $\A^\pm_\mu$ fields out of the vacuum. The
answer is $e^\al = \pm\sqrt 2\delta_{a, 28}$.  This shows that the
constant of proportionality between $e^\al$ and $\alpha^a$ is unity, i.e.
\be \label{eaaa}
e^\al = \alpha^a.
\ee

String theory also contains magnetically charged soliton states.
The magnetic charge of such a state is characterized by a vector
$Q_{mag}^\al$ defined through the equation
\be \label{defqma}
\tF^\al_{0r} \simeq {Q_{mag}^\al\over r^2} \quad \quad \hbox{for large
$r$.}
\ee
The electric and magnetic charges of a generic state are
characterized by the pair of 28 dimensional vectors $(Q_{el}^\al,
Q_{mag}^\al)$. Since elementary string states do not carry any
magnetic charge, we see that they are characterized as
\be \label{elemexit}
(Q_{el}^\al, Q_{mag}^\al) = ({1\over \lambda_2^\zero} M^\zero_{ab}
\alpha^b, 0).
\ee

Let us now consider a generic state carrying both electric
and magnetic charges. By analyzing the system containing a pair
of particles, one corresponding to an elementary string
excitation carrying charges given in Eq.\refb{elemexit}, and the
other, a generic solitonic state carrying charges $(Q_{el}^\al,
Q_{mag}^\al)$, and taking into account the
non-standard form of the gauge field kinetic term given in
\refb{2.1}, we get the following form of the
Dirac-Schwinger-Zwanziger\cite{DSZ} quantization rule,
\be\label{dirac}
\lambda_2^\zero Q_{mag}^\al (L M^\zero L)_{ab} {1\over
\lambda_2^\zero} M^\zero_{bc} \alpha^c = \hbox{integer}.
\ee
The most general solution of this equation is
\be \label{solution}
Q_{mag}^\al = L_{ab}\beta^b, \quad \quad \vec\beta\in \Lambda,
\ee
where $\Lambda$ is the self-dual Lorenzian lattice introduced
before.

We now ask the question, `what are the allowed values of
$Q_{el}^\al$ for a given $Q_{mag}^\al$?' Naively one might think that
$Q_{el}^\al$ is given by Eq.\refb{elemexit} irrespective of the
value of $Q_{mag}^\al$, but this is not the case.
{}From the analysis of Ref.\cite{WITTEN} we know that the quantization
laws for electric charge get modified in the presence of a
magnetic charge. For standard normalization of the gauge field
kinetic term, the shift is proportional to the magnetic
charge, and also the $\theta$ angle, which, in this case, is
equal to $2\pi\lambda_1^\zero$. Taking into account the
non-standard normalization of the kinetic term, and calculating
the overall normalization factor using the method of
Ref.\cite{WITTEN} (see also Ref.\cite{COLEMAN}), we get the following
spectrum of electric and magnetic charges,
\be\label{solitonexit}
(Q_{el}^\al, Q_{mag}^\al) = ({1\over \lambda_2^\zero} M^\zero_{ab}
(\alpha^b + \lambda_1^\zero \beta^b), L_{ab}\beta^b).
\ee

We now want to test if this spectrum is invariant under the SL(2,Z)
transformation given in Eq.\refb{1.10} with $a$, $b$, $c$, $d$ integers.
To test the SL(2,Z) invariance of this spectrum, we need to calculate the
transformation laws of $Q_{el}^\al$ and $Q_{mag}^\al$.  This is
straightforward, since both $Q_{el}^\al$ and $Q_{mag}^\al$ are given in
terms of the asymptotic values of the field strength $F_{\mu\nu}^\al$,
whose transformation laws are already given in Eq.\refb{1.10}. We get,
\ben \label{qeqmtrs}
Q_{el}^\al\to Q_{el}^{\al\prime} &=& (c\lambda_1^\zero +d) Q_{el}^\al + c
\lambda_2^\zero (M^\zero L)_{ab} Q_{mag}^\bet \nonumber \\
&=& {1\over \lambda_2^{\prime\zero}} M^\zero_{ab} (\alpha^{\prime b}
+ \lambda_1^{\prime \zero} \beta^{\prime b}), \nonumber \\
Q_{mag}^\al\to Q_{mag}^{\al\prime} &=& (c\lambda_1^\zero +d) Q_{mag}^\al - c
\lambda_2^\zero (M^\zero L)_{ab} Q_{el}^\bet \nonumber \\
&=& {1\over \lambda_2^{\prime\zero}} L_{ab} \beta^{\prime b},
\een
where,
\be\label{defalbepr}
\pmatrix{\vec \alpha' \cr \vec \beta'\cr} = \pmatrix{ a & -b\cr -c
& d\cr} \pmatrix{\vec \alpha \cr \vec \beta\cr}
=\LL \omega \LL^T \pmatrix{\vec \alpha \cr \vec \beta\cr},
\ee
and $\omega$ and $\LL$ have been defined in Eqs.\refb{defomega}
and \refb{defcm} respectively.
Since $a, b, c, d$ are all integers, both $\vec\alpha'$ and
$\vec\beta'$ belong to the lattice $\Lambda$. This, in turn,
shows that the $(Q_{el}^{\al\prime}, Q_{mag}^{\al\prime})$, when
expressed in terms of the transformed variables, have exactly
the same form as $(Q_{el}^\al, Q_{mag}^\al)$ before the transformation.
Hence the allowed spectrum of electric and magnetic charges in
the theory is indeed invariant under the SL(2,Z) transformation. The
crucial ingredient in this proof is that $\vec\alpha$ and $\vec\beta$
belong to the same lattice $\Lambda$, which, in turn, follows from the
fact that the lattice $\Lambda$ is self-dual.

Note that the charge spectrum that we have found refers to the charge
spectrum of all states in the theory, and not just the single particle
states. Whereas invariance of this charge spectrum under SL(2,Z)
transformation is a necessary condition for the SL(2,Z) invariance of the
theory, it is, by no means, sufficient. In order to establish the SL(2,Z)
invariance of the spectrum, we need to calculate the degeneracy
$N(\vec\alpha, \vec\beta, m)$ of single particle states of mass $m$,
characterized by charge vectors $(\vec\alpha, \vec\beta)$, and show that
it is invariant under the SL(2,Z) transformation. In particular, given any
elementary string excitation, we must be able to identify its SL(2,Z)
transforms with specific monopole and dyon states in the theory, carrying
the same mass as the elementary string state. This will be the subject of
our analysis in \S\ref{s3.3} and \S\ref{snew}.

Before we conclude this subsection, we note that under the
O(6,22) transformation given in Eq.\refb{1.3a},
\be \label{oddtrsqeqm}
Q_{el}^\al \to \Omega_{ab} Q_{el}^\bet, \quad \quad Q_{mag}^\al \to
\Omega_{ab} Q_{mag}^\bet, \quad \quad M^\zero \to \Omega M^\zero \Omega^T
\ee
which gives
\be \label{oddtrsalbe}
\pmatrix{\alpha^a \cr \beta^a\cr} \to \pmatrix{(L \Omega L)_{ab}
\alpha^b \cr (L\Omega L)_{ab} \beta^b\cr}.
\ee
Thus the charge spectrum is invariant under the O(6,22) transformation
$\Omega$ if $L\Omega L$ preserves the lattice $\Lambda$. It can be shown
that the group of such matrices form an O(6,22;Z) subgroup of
O(6,22)\cite{GIVEON}.  This establishes O(6,22;Z) invariance of the charge
spectrum.

\section{Symmetry of the Mass Spectrum}\label{s3.3}

If the string theory under consideration really has an SL(2,Z)
symmetry, then not only the allowed spectrum of electric and
magnetic charges, but also the full mass spectrum of the theory must be
invariant under the SL(2,Z) transformation. However, unlike the
spectrum of electric and magnetic charges, the mass spectrum of
the theory does receive non-trivial quantum corrections, and
hence we cannot test the SL(2,Z) invariance of the full mass
spectrum with the help of the perturbative techniques available
to us today. However, there is a special class of states in the
theory whose masses do not receive any quantum corrections.
These are the states that belong to the 16 dimensional representation of
the $N=4$ super-algebra, are annihilated by half of the sixteen
supersymmetry generators of the theory, and satisfy a definite
relation between mass and charge, known as the Bogomol'nyi
bound\cite{WITTENOLIVE}. In fact, 16-component supermultiplets exist
only for states with this special relation between mass
and charge.  Since quantum corrections cannot change the
representation to which a given supermultiplet belongs, it
cannot change the mass-charge relation of the corresponding
states either. As a result, the masses of these states do not
receive any quantum corrections\cite{WITTENOLIVE}.

Thus a consistency test of the postulate of SL(2,Z) invariance
of the theory would be to check whether the mass spectrum of the
states saturating the Bogomol'nyi bound remains invariant under
the SL(2,Z) transformation. The relationship between mass and
charge for such states can be calculated using standard
techniques\cite{HULL}. It turns out that in this case, the relevant
charges that determine the mass are the ones that also determine the
asymptotic value of the field\cite{HARLIU}
\be \label{harev1}
T_{(m)\mu\nu} \equiv \p_\mu G^\ten_{(m+3) \nu} - \p_\nu
G^\ten_{(m+3) \mu} - H^\ten_{(m+3) \mu\nu}.
\ee
Let $\wt T_{(m)}^{\mu\nu} \equiv {1\over 2}(\sqrt{-  g})^{-1} \,
\epsilon^{\mu\nu\rho\sigma} T_{(m)\rho\sigma}$, and let us stick
to the convention that all indices are raised and lowered with
the canonical metric $g_{\mu\nu}$. We now define charges $Q_m$
and $P_m$ through the asymptotic values of the fields
$T_{(m)0r}$ and $\wt T_{(m)0r}$:
\be \label{harev2}
T_{(m)0r}\simeq {Q_m\over r^2}, \quad \quad \wt T_{(m)0r} \simeq
{P_m\over r^2}.
\ee
In the normalization convention that we have been using,
the mass $m$ of a particle saturating the Bogomol'nyi bound is
determined by the following formula\cite{HARLIU}
\be \label{harev3}
m^2 = {1\over 64} \lambda_2^\zero \big( \wh G^{\zero mn} Q_m Q_n
+ \wh G^{\zero mn} P_m P_n\big),
\ee
where the matrix $\wh G_{mn}$ and its inverse $\wh G^{mn}$ have
been defined in Eq.\refb{1.2}, and the superscript $\zero$
denotes the asymptotic value as usual. Using Eqs.\refb{1.2},
\refb{defqea} and \refb{defqma} we can express $Q_m$ and $P_m$, and
hence $m^2$, in terms of $Q^\al_{el}$ and $Q^\al_{mag}$. The final
answer is\cite{SENBOGOM}
\be\label{older}
m^2 = {\lambda_2^\zero\over 16} \Big(Q_{el}^\al (LM^\zero L + L)_{ab}
Q_{el}^\bet + Q_{mag}^\al (L M^\zero L +L)_{ab} Q_{mag}^\bet\Big),
\ee
which, with the help of Eq.\refb{solitonexit} may be written
as\cite{SCHSEN2}
\be\label{bogomfor}
m^2 = {1\over 16}\pmatrix{\alpha^a & \beta^a\cr} \M^\zero (M^\zero+L)_{ab}
\pmatrix{\alpha^b\cr \beta^b\cr}.
\ee
The right hand side of this expression is manifestly invariant
under the O(6,22;Z) transformation given in \refb{1.3a} and
\refb{oddtrsalbe}, and the SL(2,Z) transformations given in
\refb{1.15} and \refb{defalbepr}\cite{SENBOGOM,SCHSEN2,ORTIN}.

This shows that two states saturating the Bogomol'nyi bound have the same
mass if their electric
and magnetic charge quantum numbers, and the asymptotic values
of moduli fields $M$ and $\lambda$, are related by an SL(2,Z)
transformation. This does not completely establish the SL(2,Z)
invariance of the mass spectrum for such states,
but shows that if the degeneracy
$N_{16}(\vec\alpha, \vec\beta)$ of 16-component supermultiplets, saturating
the Bogomol'nyi bound and carrying charge vectors $\pmatrix{\vec\alpha \cr
\vec\beta}$, is SL(2,Z) invariant, then the mass spectrum of such
states will also automatically be SL(2,Z) invariant. We shall analyze this
question in \S\ref{snew}. In particular, we shall identify the spectrum of
elementary string excitations saturating the Bogomol'nyi bound, and show
that for at least a subclass of these states, the dual magnetically
charged states are in one to one correspondence to the elementary string
excitations.

The result of this and the previous section indicates that it is more
natural to combine the two vectors $\vec\alpha$ and $\vec\beta$ into a
single 56 component vector $\pmatrix{\vec\alpha\cr \vec\beta\cr}$. This
vector belongs to a 56 dimensional lattice $\Gamma=\Lambda \otimes
\Lambda$.

\section{Symmetry of the Yukawa Couplings}\label{s5}

If SL(2,Z) is a symmetry of the theory, then all
correlation functions of the theory must be invariant under the
SL(2,Z) transformation.  In particular, various Yukawa
couplings, which represent the three point coupling between a
zero momentum scalar and two fermions (and are related to
various other couplings in the theory due to the $N=4$
supersymmetry) must also be invariant under the SL(2,Z)
transformation. However, as in the case of mass spectrum, this
symmetry can be checked only for those sets of Yukawa couplings
which do not receive any quantum corrections, {\it i.e.}, for which the
tree level answer is the exact answer. Fortunately, such Yukawa couplings
do exist in the
theory under consideration,
and, as we shall see, they are indeed invariant under the
SL(2,Z) transformation. Analysis of these Yukawa couplings will
be the subject of discussion of this section.

The Yukawa couplings under consideration are those between the
massless scalars in the theory, corresponding to the fields $M$
and $\lambda$ (or, equivalently, $\M$), and massive charged fermions
saturating the Bogomol'nyi bound. The reason that these Yukawa couplings
are given by their tree level answer is that they can be related to the
mass spectrum of the fermions, which is given by the tree level answer.
This also indicates that these Yukawa coupling must be invariant under the
SL(2,Z) (and O(6,22;Z)) transformation, since the fermion mass spectrum
has this invariance. We shall now see in some detail how this happens.

Let $M^\zero$ and $\M^\zero$ be the vacuum expectation values of
the fields $M$ and $\M$ respectively. We now introduce
fluctuations $\Phi$ and $\phi$ of these fields through the
relations
\be \label{fluct}
M = M^\zero + \Phi, \quad \quad \M = \M^\zero + \phi,
\ee
where $\Phi$ and $\phi$ are $28\times 28$ and $2\times 2$ matrices
respectively, satisfying,
\ben \label{phireln}
\Phi^T = \Phi, & \quad & \Phi L M^\zero + M^\zero L \Phi + \Phi L \Phi =0
\nonumber \\
\phi^T = \phi, & \quad & \phi \LL \M^\zero + \M^\zero \LL \phi + \phi
\LL \phi =0.
\een
The O(6,22;Z) and SL(2,Z)
transformation properties of the fields $\Phi$ and $\phi$
are given by
\be \label{phisltz}
\Phi \to \Omega \Phi \Omega^T \quad \quad \phi\to \omega \phi \omega^T,
\ee
respectively. The quanta of the fields $\Phi$ and $\phi$
are characterized by `polarization tensors' $E_{ab}$ and
$e_{\alpha\beta}$, which are symmetric $28\times 28$ and $2\times
2$ matrices respectively, satisfying,
\be \label{sltzpol}
E L M^\zero + M^\zero L E = 0, \quad \quad e \LL \M^\zero + \M^\zero
\LL e =0.
\ee
The Yukawa couplings between the $\Phi$ or $\phi$ quanta, and
the fermion fields saturating the Bogomol'nyi bound, may
now be calculated by operating $E_{ab}{\delta\over \delta
M^\zero_{ab}}$ and $e_{\alpha\beta}{\delta\over
\delta \M^\zero_{\alpha\beta}}$ on the fermion mass matrix. This
gives the following Yukawa coupling $C$ and $\wt C$ between the
fermions characterized by the electric and magnetic charge
vectors $\pmatrix{\vec \alpha\cr
\vec \beta}$ and $\pmatrix{\vec\gamma \cr \vec\delta}$, and the
scalar fields $\Phi$ and $\phi$ characterized by polarization
vectors $E$ and $e$ respectively:
\ben \label{yukawa}
C\Big( \pmatrix{\vec\alpha \cr \vec\beta}, \pmatrix{\vec\gamma \cr
\vec \delta}, E \Big) &=& {1\over 16} \pmatrix{\gamma^{a} &
\delta^{a}} \M^\zero E_{ab} \pmatrix{\alpha^b\cr \beta^b}
\times {1\over 2m(\vec\alpha, \vec\beta)
} \delta_{\vec\alpha, \vec\gamma} \delta_{\vec\beta,
\vec\delta}
\nonumber \\
\wt C\Big( \pmatrix{\vec\alpha \cr \vec\beta}, \pmatrix{\vec\gamma \cr
\vec \delta}, e \Big) &=& {1\over 16} \pmatrix{\gamma^{a} &
\delta^{a}} e (M^\zero+L)_{ab} \pmatrix{\alpha^b\cr \beta^b}
\times {1\over 2m(\vec\alpha, \vec\beta)
} \delta_{\vec\alpha, \vec\gamma} \delta_{\vec\beta,
\vec\delta}.
\een
These couplings are clearly invariant under the SL(2,Z)$\times$O(6,22;Z)
transformations:
\ben \label{sltzinvyuk}
\pmatrix{\alpha^a \cr \beta^a} \to \LL\omega\LL^T
\pmatrix{(L\Omega L)_{ab} \alpha^b
\cr (L\Omega L)_{ab}\beta^b}, & \quad &
\pmatrix{\gamma^{a} \cr \delta^{a}} \to \LL\omega\LL^T
\pmatrix{(L\Omega L)_{ab} \gamma^{b}
\cr (L\Omega L)_{ab}\delta^{b}}, \nonumber \\
E \to \Omega E \Omega^T , & \quad & e \to
\omega e \omega^T,
\een
together with the transformations of the background
\be \label{backtrs}
M^\zero \to \Omega M^\zero \Omega^T, \quad \M^\zero \to \omega \M^\zero
\omega^T.
\ee
This shows that the Yukawa couplings in a given background are equal to
the Yukawa couplings around a new background, related to the original
background by an SL(2,Z) (or O(6,22;Z)) transformation, after
appropriate transformations on the quantum numbers of the external
states.

\section{Where are the SL(2,Z) Transform of the Elementary String
Excitations?}\label{snew}

In this section we shall first identify the elementary
excitations in string theory that saturate the Bogomol'nyi
bound, and then try to identify the magnetically charged soliton
states in the theory, related to the elementary string states
via SL(2,Z) transformations\cite{SENBOGOM}. We start with a
discussion of the spectrum of known elementary string
excitations.

\subsection{Where Do the Known Elementary String Excitations Fit
in?}\label{snew1}

The mass formula \refb{bogomfor} for $\vec\beta=0$ takes the
form:
\be \label{elemone}
m^2 = {1\over 16\lambda_2^\zero} \alpha^a (M^\zero+L)_{ab} \alpha^b.
\ee
In order to compare the above formula with the mass formula for the
elementary excitations in string theory, we use the observation of
Refs.\cite{NARAIN,NSW} that the physics remains unchanged under a
simultaneous rotation of the background $M^\zero$ and the lattice
$\Lambda$ of the form:
\be \label{simrot}
M^\zero \to \Omega M^\zero \Omega^T, \quad \quad \Lambda \to
L \Omega L \Lambda.
\ee
where $\Omega$ is an O(6,22) matrix.  Certainly the mass formula is
invariant under this transformation.  If we choose $\Omega$ in such a way
that $\Omega M^\zero \Omega^T\equiv \wh M^\zero =I_{28}$, and if
$\wh{\vec\alpha}\equiv L\Omega L \vec\alpha$ denotes the vector in the new
lattice $\wh\Lambda
\equiv L\Omega L\Lambda $, then Eq.\refb{elemone} takes the simple form:
\be \label{elemtwo}
m^2 = {1\over 16\lambda^\zero_2} \wh\alpha^a (I+L)_{ab} \wh\alpha^b
={1\over 8\lambda_2^\zero} (\wh{\vec\alpha}_R)^2,
\ee
where,
\be \label{elemthree}
\wh{\alpha}^a_R \equiv {1\over 2} (I+L)_{ab} \wh\alpha^b, \quad
\quad
\wh{\alpha}^a_L \equiv {1\over 2} (I-L)_{ab} \wh\alpha^b.
\ee

We now write down the mass formula for elementary string
excitations\cite{GROSS}.
Since the Ramond sector states are degenerate with the Neveu-Schwarz (NS)
sector states due to space-time supersymmetry, it is enough to study the
mass formula in the NS sector.
With the normalization that we have chosen, it is given by
\be \label{elemfour}
M^2 = {1\over 8\lambda_2^\zero} \{ (\wh{\vec\alpha}_R)^2 + 2 N_R -1\}
= {1\over 8\lambda_2^\zero} \{ (\wh{\vec\alpha}_L)^2 + 2 N_L -2\}.
\ee
In this expression $(\wh{\vec\alpha}_R)^2$ and $(\wh{\vec\alpha}_L)^2$
denote the internal momenta contributions, $N_L$ and $N_R$ denote the
oscillator contributions, and $-1$ and $-2$ denote the ghost contributions
to $L_0$ and $\bar L_0$ in the world-sheet theory respectively.
Note that in our convention the world-sheet supersymmetry appears in
the right moving sector of the theory.
The appearance of ${1/\lambda_2^\zero}$ factor in these
expressions can be traced to the fact that we are using the canonical
metric $g_{\mu\nu}$ to measure distances instead of the string metric
$G_{\mu\nu}$. GSO projection requires $N_R$ to be at least $1/2$, since
we need a factor of $\psi^M_{-1/2}$ to create the lowest mass state in
the NS sector. This clearly shows that $M^2\ge m^2$ with $m^2$ given by
Eq.\refb{elemtwo}. Furthermore the elementary string states that saturate
the Bogomol'nyi bound all have
\be \label{elemfive}
N_R = {1\over 2}
\ee
so that
\be \label{elemfoura}
M^2 = {1\over 8\lambda_2^\zero} \{ (\wh{\vec\alpha}_R)^2 \}
= {1\over 8\lambda_2^\zero} \{ (\wh{\vec\alpha}_L)^2 + 2 N_L -2\}.
\ee
For these states, $M^2=m^2$. We also see that
\be\label{nl1}
N_L-1 = {1\over 2} \big( (\wh{\vec\alpha}_R)^2 - (\wh{\vec\alpha}_L)^2
\big) = {1\over 2} \wh\alpha^a L_{ab} \wh\alpha^b \equiv {1\over 2}
(\wh{\vec\alpha})^2 .
\ee
Since space-time supersymmetry generators act
only on the right-moving fermions $\psi^M$, it is also easy to
analyze the supersymmetry transformation properties of these states. In
particular, for a fixed oscillator state in the left-moving sector, states
created by $\psi^M_{-1/2}$ for eight transverse $M$, together with their
Ramond sector counterparts, give rise to a 16 dimensional super-multiplet
of the N=4 supersymmetry algebra. The transformation laws of these states
under the full N=4 super-Poincare algebra, however, depend on the left
moving oscillator content also. In particular, if the left moving
oscillators involved in the construction of a state transform as a scalar,
then the resulting supermultiplet will contain states with maximum spin 1,
we shall call this the vector supermultiplet. On the other hand, if the
left-moving oscillators transform as a vector, then the resulting
supermultiplet contains states with maximum spin 2. We shall refer to this
representation of the super-Poincare algebra as the spin 2 supermultiplet.
It should be clear from this discussion that super-multiplets of
arbitrarily high spin can be constructed this way. However, each of these
super-multiplets decompose into several copies of the 16 dimensional
super-multiplet if we look at their transformation laws under the
supersymmetry subalgebra of the full super-Poincare algebra.

Before we conclude this subsection, let us analyse the stability of the
various elementary string excitations discussed above.\footnote{ I wish to
thank A. Strominger for raising this issue.}  Since we are concentrating
on states saturating the Bogomol'nyi bound, we are guaranteed that in the
rest frame these states are the lowest energy states in the given charge
sector, and hence there is no multiparticle state in theory that carries
the same amount of charge and has {\it less} energy than the single
particle state. It is, however, possible that there exists a multiparticle
state, with all the particles at rest, which has the {\it same} energy as
the particular elementary particle state under consideration. {}From the
mass relation $m^2\propto (\wh{\vec\alpha}_R)^2$ it is clear that such a
situation can arise if the right hand component of the charge vector
$(\wh{\vec\alpha}_R)$ of the original state, and those of the states
constituting the multi-particle state, are parallel to each other.  If
such a situation has to hold for a generic choice of the lattice, then it
would imply that the full charge vectors $\wh{\vec\alpha}$ of the original
particle, and of the decay products, must also be parallel to each other.
For, if the right hand components of the charge vectors are parallel to
each other but the left hand components are not, then a slight O(6,22)
rotation of the lattice, which mixes the right and the left hand
components of the charge vectors, will destroy the alignment of the right
hand components.  This implies that in order for a particle carrying
charge vector $\wh{\vec\alpha}$ to decay into two or more particles at
rest, $\wh{\vec\alpha}$ must be an integral ($n$) multiple of some other
lattice vector $\wh{\vec\alpha}_0$. In this case the original particle can
decay into $n$ other particles, each carrying charge vector
$\wh{\vec\alpha}_0$. {}From this we can conclude that for a
generic choice of
the lattice, an elementary string state, saturating the Bogomol'nyi bound,
and characterized by the charge vector $\wh{\vec\alpha}$, is absolutely
stable as long as $\wh{\vec\alpha}$ is not an integral multiple of another
vector in the lattice $\wh\Lambda$.

In \S\ref{snew3} we shall try to identify soliton states of the
theory which are related to these elementary string excitations
via SL(2,Z) transformations. But first we need to know how the
soliton solutions in the theory fit into the mass formula given in
Eq.\refb{bogomfor}.

\subsection{Where Do the Known Solitons Fit
in?}\label{snew2}

We now turn our attention to the spectrum of known magnetically
charged soliton solutions in string theory. Many such solutions
are known\cite{BANKS,HARLIU,KHURI,GHL}. We
shall focus our attention only on the non-singular solutions
with asymptotically flat space-time geometry, since it is only these
solutions which have a clear interpretation as new particle like
states in the theory.

{\bf BPS Gauge Monopole Solutions}: These solutions were
constructed in Ref.\cite{HARLIU} (see also Ref.\cite{BANKS}) and were
further explored in Ref.\cite{GHL}. We work in a gauge where asymptotically
the Higgs field is directed along a fixed direction in the gauge space
(and is identified with the field $A^{\ten 16}_4$) except along a Dirac
string singularity. In this gauge, after appropriate rescaling of the ten
dimensional coordinates $z^0$ and $z^4$,
the asymptotic values of various
ten dimensional fields associated with this solution are given by
\ben\label{tendmonopole}
&& B^\ten_{\mu\nu}\simeq 0, \quad \quad  G^\ten_{\mu\nu} \simeq
e^{2\phi_0}\eta_{\mu\nu}, \quad \quad \Phi^\ten \simeq 2\phi_0,
\nonumber \\
&& G^\ten_{(m+3)\mu} = 0, \quad \quad H^\ten_{(m+3)0i}\simeq O({1\over
r^3}), \quad \quad
H^\ten_{(m+3)ij} \simeq 8C e^{-\phi_0}\delta_{m,1} \varepsilon_{ijk}
{x^k\over r^3}, \nonumber \\
&& F^{\ten I}_{0i}\simeq O({1\over r^3}), \quad \quad F^{\ten I}_{ij} \simeq
-2\sqrt 2 \, \delta_{I,16} \varepsilon_{ijk} {x^k\over r^3}, \nonumber \\
&& B^\ten_{(m+3)(n+3)} \simeq 0, \quad \quad A^{\ten I}_{m+3} \simeq
2\sqrt 2 C e^{-\phi_0} \delta_{I,16} \delta_{m,1}, \quad \quad
G^\ten_{(m+3)(n+3)} \simeq \delta_{mn}, \nonumber \\
&& \quad \quad 1\le i,j\le 3, \quad 1\le m,n \le 6, \quad 0\le \mu, \nu
\le 3,
\een
where $C$ and $\phi_0$ are two arbitrary constants. Using Eqs.\refb{1.2}
we see that the asymptotic values of various four dimensional fields are
given by,
\ben \label{monopole1}
\wh G_{mn} & \simeq & \delta_{mn}, \quad\quad \wh B_{mn} \simeq 0,
\quad\quad \wh A^I_m \simeq 2\sqrt 2\, C e^{-\phi_0} \delta_{I,16}
\delta_{m,1}, \quad\quad \Phi = 2\phi_0, \nonumber \\
F^\al_{0r} & \simeq & O({1\over r^3}), \quad \quad \wt F^\al_{0r} \simeq
- \sqrt 2 \,
\delta_{a, 28}\, {1 \over r^2}, \quad\quad
g_{\mu\nu}\simeq \eta_{\mu\nu}, \quad\quad B_{\mu\nu}\simeq 0,
\een
Note that even though $H^\ten_{(m+3)ij}$ is asymptotically non-trivial,
$F^{(m+6)}_{ij}$ is trivial. This happens due to the cancellation between
various terms appearing in the expression for $A^{(m+6)}_\mu$ given in
Eq.\refb{1.2}.

This solution can be generalized in
several ways. In particular, we can generate a multi-parameter family of
solutions, if, {\it keeping the lattice $\Lambda$ fixed}, we make the
following transformations on the original solution:
\ben \label{newsoln1}
&& G^\ten_{(m+3)(n+3)} \to S_m^{~p} S_n^{~q} G^\ten_{(p+3)(q+3)}, \quad
\quad A^{\ten I}_{(m+3)} \to S_m^{~p} A^{\ten I}_{p+3} + T^I_m,
\nonumber \\
&& B^\ten_{(m+3)(n+3)} \to S_m^{~p} S_n^{~q} B^\ten_{(p+3)(q+3)} + R_{mn}
+{1\over 2} (S_m^{~p} A^{\ten I}_{p+3} T^I_n - S_n^{~p} A^{\ten I}_{p+3}
T^I_m), \nonumber \\
&& G^\ten_{(m+3)\mu} \to S_m^{~p} G^\ten_{(p+3)\mu}, \quad \quad
B^\ten_{(m+3)\mu} \to S_m^{~p} B^\ten_{(p+3)\mu} - {1\over 2} A^{\ten
I}_\mu T^I_m,
\een
where $S_m^{~p}$ is an arbitrary constant $6\times 6$ matrix, $R_{mn}$ is
a constant anti-symmetric $6\times 6$ matrix, and $T^I_m$ is a constant
$6\times 16$ matrix, satisfying,
\be \label{newsoln2}
T^{16}_m=0.
\ee
All other 10 dimensional fields remain invariant under these
transformations. The freedom of shifting $B^\ten_{(m+3)(n+3)}$ and
$A^{\ten I}_{(m+3)}$ by constant matrices $R_{mn}$ and $T^I_m$ stem from
the fact that the equations of motion involve only the field strengths
$H^\ten_{MNP}$ and $F^{\ten I}_{MN}$. These field strengths are invariant
under these transformations, as can be seen from Eqs.\refb{new2}. The
reason that $T^{16}_m$ need to vanish is that the solution contains
SU(2) gauge fields $\A^{\ten i}_M$ ($1\le i\le 3$) at its core, with
$A^{\ten 16}_M$ identified to $2\sqrt 2\,\A^{\ten 3}_M$.  Thus a constant
shift in $A^{\ten 16}_M$ will change the SU(2) field strengths, and the
resulting configuration will not remain a solution of the equations of
motion.

Performing the transformations \refb{newsoln1} on the solution
\refb{tendmonopole}, and using Eqs.\refb{1.2} again, we get the following
asymptotic form of various four dimensional fields,
\ben \label{newsoln3}
\wh G_{mn} & \simeq & S_m^{~p} S^{~p}_n, \quad\quad \wh B_{mn} \simeq
R_{mn},
\quad\quad \wh A^I_m \simeq 2\sqrt 2 \, C e^{-\phi_0} \delta_{I,16}
S_m^{~1}  + T_m^I, \quad\quad \Phi = 2\phi_0 - \det S,
\nonumber \\
F^\al_{0r} & \simeq & O({1\over r^3}), \quad \quad \wt F^\al_{0r} \simeq
- \sqrt 2 \,
\delta_{a, 28}\, {1 \over r^2}, \quad\quad
g_{\mu\nu}\simeq \eta_{\mu\nu}, \quad\quad B_{\mu\nu}\simeq 0.
\een
It can be checked that by appropriately adjusting the matrices $S$, $T$
and $R$, and the constant $C$, we can choose $\wh G^\zero_{mn}$, $\wh
B^\zero_{mn}$ and $\wh A^{\zero I}_m$ to be completely arbitrary,
consistent with their symmetry properties. Thus the monopole solution
given in Eq.\refb{newsoln3} is characterized by an arbitrary value of
$M^\zero$.

Using Eqs.\refb{defqea}, \refb{defqma}, and
\refb{solitonexit} we see that this monopole carries quantum
numbers
\be \label{monopole2}
\pmatrix{\alpha^a = 0 \cr \beta^a = \sqrt 2 \delta_{a,28}},
\ee
with
\be \label{monopole3}
\lambda^\zero_1 =0.
\ee
The BPS dyon solutions, saturating the Bogomol'nyi bound were
also constructed in Ref.\cite{HARLIU} following the method of
Ref.\cite{DADDA}. For these solutions,
\be \label{monopole4}
F^\al_{0r} \simeq \sqrt 2 \, {Q_e\over \lambda_2^\zero} M^\zero_{a, 28} {1
\over r^2},
\ee
instead of being zero. Here $Q_e$ is an arbitrary constant. Using
Eqs.\refb{defqea} and
\refb{solitonexit}, it is easy to see that these
solutions correspond to non-zero values of $\lambda_1^\zero$ and
carry quantum numbers,
\be \label{monopole5}
\pmatrix{\alpha^a = p \sqrt 2 \delta_{a,28} \cr \beta^a = \sqrt
2 \delta_{a,28}},
\ee
with $\lambda_1^\zero$ and the integer $p$ determined (up to the SL(2,Z)
transformation $\lambda_1^\zero \to \lambda_1^\zero - n$, $p\to p+n$ for
some integer $n$) in terms of the parameter $Q_e$ by the relation $Q_e =
p+\lambda_1^\zero$. Following the arguments of Ref.\cite{OSBORN} one can
show that these states belong to the vector supermultiplet of the
super-Poincare algebra.

In the next subsection we shall compare these states with the SL(2,Z)
transform of the elementary string excitations discussed in the last
subsection. Note, however, that the analysis of the last subsection was
carried out in a representation where the matrix $M^\zero$ was transformed
to the identity matrix via an O(6,22) rotation, and all the modular
parameters were encoded in the lattice $\wh\Lambda$. In order to
facilitate comparison, it is convenient to bring $M^\zero$ to identity in
this case also, with a simultaneous rotation of the lattice $\Lambda$ to
$\wh\Lambda$. Under this rotation, the vector $\sqrt 2\, \delta_{a,28}$ is
transformed to some vector ${\vec l}$ with ${\vec l}^2 \equiv l^a
L_{ab} l^b =-2$. Thus the resulting dyon solution has charge quantum
numbers
\be\label{dyonnew}
\pmatrix{\wh{\vec\alpha} = p \vec l \cr \wh{\vec\beta} = \vec l\cr}.
\ee
Applying this argument in reverse, we can construct dyon solutions at any
point in the moduli space, characterized by some self-dual Lorentzian
lattice $\wh\Lambda$, with $\wh M^\zero=I_{28}$. For, given such a
configuration, we can always find an O(6,22) transformation $\Omega$ such
that $\wh\Lambda=L\Omega L\Lambda$. This transformation rotates $\wh
M^\zero=I_{28}$ to $M^\zero=(\Omega^T\Omega)^{-1}$.  This rotation also
brings some vector $\vec l\in\wh\Lambda$ with $\vec l^2=-2$ to the vector
$\sqrt 2\, \delta_{a,28}\in \Lambda$. Since for the compactification lattice
$\Lambda$ we know how to construct a dyon solution with charge vector
\refb{monopole5} for any value of $M^\zero$, the O(6,22) rotation of this
solution by $\Omega$ will give us a dyon solution carrying charge quantum
numbers \refb{dyonnew} in the vacuum characterized by the lattice
$\wh\Lambda$ and $\wh M^\zero =I_{28}$. Also, note that the transformation
$\Omega$ that gives $\wh\Lambda=L\Omega L\Lambda$ is not unique,
since $L\Omega L$ can always be multiplied from the right by any element
of the O(6,22;Z) subgroup of O(6,22) that constitutes the group of
automorphisms of the lattice $\Lambda$. Using this freedom, different
vectors $\vec l\in \wh\Lambda$ can be mapped to the vector $\sqrt 2\,
\delta_{a,28}\in \Lambda$. This gives us a way of constructing dyon
solutions carrying charge quantum numbers \refb{dyonnew} for different
vectors $\vec l\in \wh\Lambda$ with $\vec l^2=-2$.

We should note, however, that the solutions of Ref.\cite{HARLIU} were
constructed by ignoring the higher derivative terms in the string
effective action, and hence are valid for small $C$, which in this case
translates to small $({\vec l_R})^2$.\footnote{To see this, note that
small $C$ with the standard choice of the lattice $\Lambda$ implies
small mass for the particles carrying charge quantum numbers $\pm \sqrt
2\, \delta_{a,28}$, $-$ these particles can be interpreted as the SU(2)
gauge bosons that have acquired mass due to spontaneous breakdown of the
SU(2) symmetry by the Higgs vacuum expectation value $\propto C$. On the
other hand, in the picture where $M^\zero$ has been set to identity by an
O(6,22) rotation, Eq.\refb{elemfour} tells us that for $N_R=1/2$,
particles carrying electric charge vector $\vec l$ has mass$^2$
proportional to $(\vec l_R)^2$. This shows that small $C$ in one picture
implies small $(\vec l_R)^2$ in the other picture.}  Nevertheless we
expect that the general features of the solution, {\it e.g.}, partially
broken supersymmetry, will continue to hold for all $C$, and consequently,
it will continue to represent a state in the vector representation of the
super-Poincare algebra, saturating the Bogomol'nyi bound.

{\bf $H$-Monopole Solutions}: We now turn to the next class of
solutions in string theory, which carry magnetic charge
associated with the ten dimentional field $H^\ten_{MNP}$ but not
the ten dimensional gauge fields\cite{KHURI,GHL}. A
non-singular, asymptotically flat solution of this kind was
constructed in Ref.\cite{GHL} by wrapping {\it a finite sized
gauge five-brane solution} around the torus. After appropriate rescaling
of the ten dimensional coordinates $z^0$ and $z^4$, the only non-trivial
asymptotic fields for this solution are given by,
\be\label{anothereq}
\Phi^\ten \simeq 2\phi_0, \quad \quad G^\ten_{\mu\nu}=e^{2\phi_0}
\eta_{\mu\nu}, \quad \quad H^\ten_{(m+3)ij} \simeq 2Q\delta_{m,1}
\epsilon_{ijk} {x^k\over r^3},
\ee
where $Q$ is a constant. {}From this we can determine the asymptotic values
of various four dimensional fields. They are,
\ben \label{hmonopole1}
\wh G_{mn} & \simeq & \delta_{mn}, \quad\quad \wh B_{mn} \simeq 0,
\quad\quad \wh A^I_m \simeq 0, \quad\quad \Phi = 2\phi_0,
\nonumber \\ F^\al_{0r} & \simeq & O({1\over r^3}), \quad \quad \wt
F^\al_{0r} \simeq Q \delta_{a, 7} {1\over r^2}, \quad\quad
g_{\mu\nu}\simeq \eta_{\mu\nu}, \quad\quad B_{\mu\nu}\simeq 0.
\een
Using Eqs.\refb{newa1}, \refb{defqea}, \refb{defqma}, and
\refb{solitonexit} we see that this monopole carries quantum
numbers
\be \label{hmonopole2}
\pmatrix{\alpha^a = 0 \cr \beta^a = Q \delta_{a,1}},
\ee
with,
\be \label{hmonopole3}
\lambda^\zero_1 =0.
\ee
Since $\vec\beta$ lies on the lattice $\Lambda$, we see that the
parameter $Q$ must be quantized.
Since this solution does not carry any electric charge, the
corresponding value of $\lambda_1^\zero$ is 0. Although the
corresponding dyon solutions have not been constructed, there
is, in principle, no reason to expect that they do not exist.
These dyon solutions will correspond to non-zero values of
$\vec\alpha$ and $\lambda_1^\zero$ as before.

For the solution given in Eq.\refb{hmonopole1}, $M^\zero = I_{28}$, but as
in the previous case, we can get more general class of solutions using the
transformations \refb{newsoln1}. Since this monopole solution contains
SU(2) gauge fields at its core\cite{GHL}, the transformation parameter
$T_m^I$ must satisfy an equation similar to Eq.\refb{newsoln2}. In fact if
we take $A^{\ten 16}_M$ to be the third component $\A^{\ten 3}_M$ of the
SU(2) gauge field, then the condition on $T_m^I$ is precisely the one
given in \refb{newsoln2}. As a result, even after the transformation, we
have $\wh A^{(28)}_m=0$ asymptotically. This shows that
by this method, monopole solutions
carrying charge quantum numbers \refb{hmonopole2} cannot be constructed
for arbitrary choice of $M^\zero$, but only for a specific class of
$M^\zero$.

As in the previous case, we can bring $M^\zero$ to $I_{28}$ by an O(6,22)
rotation, simultaneously rotating the lattice $\Lambda$ to a new lattice
$\wh\Lambda$. The vector $Q\delta_{a,1}$ gets rotated into some new vector
${\vec m}$ satisfying ${\vec m}^2=0$. Thus the charge quantum
numbers of the monopole are now given by,
\be \label{newsoln4}
\pmatrix{\wh{\vec \alpha} =0 \cr \wh{\vec\beta}=\vec m}.
\ee
The fact that the $H$-monopole solutions can be constructed only for a
special class of $M^\zero$ now translates into the statement that such
solutions exist only for a special class of lattice $\wh\Lambda$, $-$
those which correspond to the existence of an unbroken SU(2) gauge group.

\subsection{SL(2,Z) Transform of the Elementary String
States}\label{snew3}

In this subsection we shall try to identify soliton solutions related to
the elementary string excitations via SL(2,Z) transformation. We begin by
reminding the reader that the SL(2,Z) transformation acts non-trivially on
the vacuum, and hence relates elementary string excitations in one vacuum
to the monopole and dyon solutions constructed around different
vacua. Throughout this discussion we shall be implicitly assuming that the
theory is in a single phase in the entire upper half $\lambda^\zero$
plane, unlike the cases discussed in Refs.\cite{CARDY,SHAPERE}, so that
the dyon spectrum computed at weak coupling can be continued to the strong
coupling regime.\footnote{ This is analogous to the fact that the theory
is in the same phase for all values of $M^\zero$, except possibly on
surfaces of high codimension in the moduli space, where part of
the non-abelian gauge symmetry is unbroken.}

We shall concentrate on the states belonging to
the 16 dimensional representation of the supersymmetry algebra. The mass
spectrum of such states has been given in Eq.\refb{elemfoura}.  We shall
discuss the three cases, $(\wh{\vec\alpha})^2 =-2$, $(\wh{\vec\alpha})^2
=0$, and $(\wh{\vec\alpha})^2 > 0$ separately.

\noindent {\bf $(\wh{\vec\alpha})^2 =-2$}:
In this case Eq.\refb{nl1} gives $N_L=0$. Since there are no left moving
oscillators, by our previous argument, these states, together with their
Ramond sector counterparts, constitute a vector supermultiplet of the
super-Poincare algebra. Note also that each of these particles are
absolutely stable, since the lattice $\Lambda$, being even and self dual,
cannot contain $\wh{\vec\alpha}/n$ as a lattice vector for any integer
$n$. Under the SL(2,Z) transformation
\be \label{elemtwelve}
\LL\omega\LL^T = \pmatrix{0 & -1 \cr 1 & 0\cr}
\ee
an elementary string state carrying charge quantum numbers
$\pmatrix{\wh{\vec \alpha} = \vec l\cr \wh{\vec \beta}=0}$ is mapped onto
a soliton state carrying charge quantum numbers given in Eq.\refb{dyonnew}
with $p=0$. Furthermore, as we have seen, these magnetically charged
states can be constructed for any choice of the vacuum characterized by
the lattice $\wh\Lambda$. This agrees with the fact that the elementary
string states of the form discussed above also exist for any choice of the
lattice $\wh\Lambda$.  Finally, as has already been mentioned before,
these soliton states belong to the vector supermultiplet of the $N=4$
super-Poincare algebra\cite{OSBORN}.  This shows that for
elementary string states saturating the Bogomol'nyi bound and having
$N_L=0$, we do have soliton states in the theory related to these
elementary string states via the SL(2,Z) transformation \refb{elemtwelve},
and belonging to the same representation of the super-Poincare algebra.

Let us now analyze the effect of a general SL(2,Z)
transformation on an elementary string state labeled by the
quantum numbers $\pmatrix{\wh\alpha^a = l^a \cr
\wh\beta^a=0}$, with $\vec l^2 =-2$. Acting on such a state, an SL(2,Z)
transformation
\be\label{elemthirteen}
\LL\omega \LL^T = \pmatrix{p & q\cr r & s\cr}, \quad ps-qr=1,
\ee
produces a state with quantum numbers
\be\label{elemfourteen}
\pmatrix{\wh\alpha^a= p l^a \cr \wh\beta^a =
r l^a }.
\ee
Note that the quantum numbers of the final state depend only on
$p$ and $r$. Given $p$ and $r$ which are relatively prime, it is
always possible to find $q$ and $s$ satisfying $ps-qr=1$.
Furthermore, the choice of $q$ and $s$ is unique up to a
translation $s\to s+nr$, $q\to q+np$ for some integer $n$. This
freedom can be understood by noting that
\be\label{elemfifteen}
\pmatrix{ p & q+np \cr r & s + nr} = \pmatrix{ p & q\cr r & s}
\pmatrix{1 & n \cr 0 & 1\cr}.
\ee
The SL(2,Z) transformation $\pmatrix{1 & n \cr 0 & 1}$, acting
on an elementary string state carrying only electric charge,
leaves its quantum numbers unchanged. Acting on the field
$\lambda$, it produces the trivial transformation $\lambda\to
\lambda-n$. Thus we see that up to this trivial transformation,
different SL(2,Z) group elements, labeled by the integers $p$
and $r$, produce different charge quantum numbers.

{}From this analysis we conclude that in order to establish SL(2,Z)
invariance of the spectrum in this sector, one needs to show the existence
of non-singular, asymptotically flat, dyon solutions carrying charge
quantum numbers given in Eq.\refb{elemfourteen} for all relatively prime
integers $p$ and $r$. Furthermore, these dyon states must saturate the
Bogomol'nyi bound and belong to the vector representation of the
super-Poincare algebra. The soliton states carrying charge quantum numbers
given in \refb{dyonnew} are special cases of these with $r=1$.

Existence of these new dyon states in the theory can be taken to be a
prediction of the SL(2,Z) invariance of the theory. Let us now give a
plausibility argument for the existence of these states. We begin with
the observation that the charge quantum numbers with $r>1$ correspond to
states with multiple units of magnetic charge. Multi-dyon solutions in
ordinary Yang-Mills-Higgs system have been constructed in the BPS
limit\cite{CORRIGAN,MULLER}, and there is good reason to believe that
they also exist in the full string theory\cite{HARLIU}. It is quite
plausible that when we quantize the bosonic and the fermionic zero modes
of these solutions, then in each charge sector, the ground state will have
partially broken supersymmetry, and will belong to the vector
supermultiplet of the super-Poincare algebra, as in the case of singly
charged monopoles. What is not so obvious is what is special about the
cases when $p$ and $r$ are relatively prime. We shall now show that dyons
carrying quantum numbers given in Eq.\refb{elemfourteen} represent
absolutely stable single particle states if and only if $p$ and $r$ are
relatively prime. These dyons could then be regarded as stable,
supersymmetric, bound states of monopoles and dyons, each carrying one
unit of magnetic charge.

Suppose $p$ and $r$ are not relatively prime, so that there
exist integers $p_0$, $r_0$ and $n$ such that $p=n p_0$ and $r=n
r_0$.  It is easy to verify that a dyon with quantum number
\be\label{elemeighteen}
\pmatrix{\wh\alpha^a =  n p_0 l^a \cr \wh\beta^a =
n r_0 l^a}
\ee
and saturating the Bogomol'nyi bound, has mass and charge
identical to that of $n$ dyons with quantum numbers
\be\label{elemnineteen}
\pmatrix{\wh\alpha^a =  p_0 l^a  \cr \wh\beta^a =
r_0 l^a}
\ee
and hence is indistinguishable from such a state. Thus these
dyons should not be regarded as new states in the spectrum. On
the other hand, if $p$ and $r$ are relatively prime, then the
dyon with charge quantum numbers given in Eq.\refb{elemfourteen}
cannot be regarded as a state containing multiple dyons, since
the mass of this dyon is strictly less than the sum of the
masses of the dyons whose charge quantum numbers add up to those
given in Eq.\refb{elemfourteen}. To see this, let us compare the
mass of the dyon with charge quantum numbers given in
\refb{elemfourteen} to the sum of the masses of the dyons carrying
charge quantum numbers
\be\label{elemtwenty}
\pmatrix{\wh\alpha^a= p_1 l^a \cr \wh\beta^a =
r_1 l^a}\quad \hbox{and} \quad
\pmatrix{\wh\alpha^a= p_2 l^a \cr \wh\beta^a =
r_2 l^a}, \quad \hbox{with} \quad p=p_1+p_2, \,\, r=r_1+r_2.
\ee
One can easily verify that the mass of the dyon carrying charge
quantum numbers given in Eq.\refb{elemfourteen} is smaller than the
sum of the masses of the dyons carrying charge quantum numbers
given in Eq.\refb{elemtwenty}, by using the triangle inequality
\be \label{elemtwentyone}
\bigg[\pmatrix{ p & r} \M^\zero \pmatrix{p\cr r}\bigg]^{1\over 2}
\le \bigg[ \pmatrix{ p_1 & r_1}
\M^\zero \pmatrix{p_1\cr r_1} \bigg]^{1\over 2}
+ \bigg[ \pmatrix{ p_2 & r_2} \M^\zero
\pmatrix{p_2\cr r_2}\bigg]^{1\over 2},
\ee
and noting that the equality holds if and only if $p_1/r_1 = p_2/r_2=
p/r$, which cannot happen if $p$ and $r$ are relatively prime. Thus for
$p$ and $r$ relatively prime, the dyons carrying quantum numbers given in
Eq.\refb{elemfourteen} are absolutely stable, and should be regarded as
new states in the theory.

\noindent {\bf $(\wh{\vec\alpha})^2 =0$}:
In this case Eq.\refb{nl1} gives $N_L=1$.  The contribution to $N_L$ here
can come from the oscillators associated with any of the 22 internal
directions, or the four space-time directions. The oscillators associated
with the 22 internal directions transform as scalars under the four
dimensional Lorentz transformation, and hence give rise to vector
super-multiplets of the super-Poincare algebra.  The requirement that the
corresponding vertex operator is a primary operator gives one constraint,
which reduces the number of independent choices of the left moving
oscillator to 21. Thus there are 21 distinct vector supermultiplets of the
super-Poincare algebra at this level. On the other hand, the left moving
oscillators associated with the space-time coordinates transform as
vectors under the four dimensional Lorentz transformation. By our previous
argument, this gives rise to a spin two supermultiplet of the
super-Poincare algebra.

Note that given any light-like vector $\wh{\vec\alpha}\in\wh\Lambda$,
$n\wh{\vec\alpha}$ is also a like-like vector in the lattice $\Lambda$.
However, the later state can decay into $n$ particles at rest, each
carrying charge vector $\wh{\vec\alpha}$.

The SL(2,Z) transformation
\refb{elemtwelve} maps elementary string states carrying charge quantum
nunbers $\pmatrix{\wh{\vec \alpha} = \vec m\cr \wh{\vec \beta}=0}$ with
$\vec m^2=0$ to monopole states carrying charge quantum numbers
$\pmatrix{\wh{\vec \alpha} = 0 \cr \wh{\vec \beta} = \vec m}$.
This coincides with the quantum numbers of the $H$-monopole
solution given in \refb{newsoln4}. However, note
that these $H$-monopole solutions have been constructed only
for a subclass of vacuum configurations, whereas the elementary
string states carrying the quantum number $\wh{\vec\alpha}=\vec m$
exist for all choices of the vacuum.

If SL(2,Z) is a genuine symmetry of the theory, then there
should be a one to one correspondence between the elementary
string states and monopole solutions of this kind, and hence one
must be able to construct the $H$-monopole solutions for a
generic choice of the lattice $\wh\Lambda$. Also, there should be 21
distinct $H$-monopole states in the vector representation and 1
$H$-monopole state in the spin 2 representation of the super-Poincare
algebra, carrying the same magnetic charge, since the elementary string
state carrying a given electric charge has this degeneracy. Finally there
should be $H$-dyon states carrying $p$ units of electric charge and $r$
units of magnetic charge for $p$ and $r$ relatively prime.  Existence of
these states can be taken to be a prediction of the SL(2,Z) invariance of
the theory. One already sees evidence of large degeneracies in the
construction of the $H$-monopole solution in Ref.\cite{GHL}, since an
SU(2) gauge group is necessary to construct the solution, and different
choices of this SU(2) group will lead to different $H$-monopole solutions
carrying the same charge quantum numbers.\footnote{ The charge quantum
numbers of the $H$-monopole are not affected by the choice of the SU(2)
group.} However, a proper understanding of this degeneracy will be
possible only after we are able to construct the $H$-monopole solution in
a generic background where the non-abelian gauge group of the theory is
completely broken, and then quantize the bosonic and fermionic zero modes
of the solution.

\noindent{\bf $(\wh{\vec\alpha})^2 > 0$}:
In this case, from Eq.\refb{nl1} we get $N_L\ge 2$.
These states carry charge quantum numbers of the form
\be \label{elemtwentysix}
\pmatrix{ \wh{\vec \alpha} = \vec n\cr
\wh{\vec \beta}=0},
\ee
with $\wh{\vec n}^2=2(N_L-1)>0$.
The monopoles, related to these states by the SL(2,Z) transformation
\refb{elemtwelve}, have quantum numbers
\be \label{elemtwentyseven}
\pmatrix{ \wh{\vec\alpha} = 0 \cr
\wh{\vec\beta} = \vec n}.
\ee
There are no known monopole solutions carrying these quantum numbers.
This, however, is not surprising, since, as we shall argue now,
there is no {\it a priori} reason why such monopole solutions can be
constructed in terms of the massless fields of the low energy effective
field theory. Note that in the previous two cases, there is a limit
($(\wh{\vec\alpha}_R)^2\to 0$) in which the monopole mass vanishes,
and hence, at least in this limit, the monopole solution must be
constructed purely in terms of the massless fields of the theory. In the
present case, however, there is no such limit since
$(\wh{\vec\alpha}_R)^2\ge 2$, and these monopoles  always have  mass of
order $M_{Pl}$. Thus there is no reason to expect that these monopoles
can be constructed in terms of the massless fields in the low energy
effective field theory. Construction of monopole solutions carrying these
quantum numbers remains another open problem in this field.

\section{SL(2,Z) Duality in String Theory as Target Space Duality of the
Five Brane Theory} \label{s4}

In the previous sections we have presented several pieces of evidence
that the SL(2,Z) symmetry, which exchanges the strong and weak
coupling limits of the string theory, is a genuine symmetry of the
theory. The purpose of this section is somewhat different;
instead of producing more evidence for the SL(2,Z) symmetry, we
shall try to find a geometrical understanding of this
symmetry.

We begin with the observation that the O(6,22;Z) symmetry
already has a nice geometrical interpretation. It generalizes
the symmetry that sends the size of the compact manifold, measured in
appropriate units, to its inverse, and, at the same time, exchanges the
usual Kaluza-Klein modes of the string theory carrying momentum in the
internal directions, with the string winding modes, $-$ states
corresponding to a string wrapped around one of the compact directions.
One way to see this is to note that the six dimensional vector $\alpha^m$
($1\le m \le 6$) has the interpretation as the components of momentum of a
state in the internal directions, and $\alpha^{m+6}$ ($1\le m\le 6$) has
the interpretation as the winding number of a state along the compact
directions. Thus the O(6,22;Z) transformation $\pmatrix{0 & I_6 & 0 \cr
I_6 & 0 & 0\cr 0 & 0 & I_{16}}$ gives $\alpha^m\leftrightarrow
\alpha^{m+6}$ for $1\le m \le 6$, thereby interchanging the quantum
numbers associated with internal momenta and winding numbers.

No such simple geometric interpretation exists for SL(2,Z)
transformation in string theory. In fact, as we have seen, the
non-trivial part of the SL(2,Z) transformation exchanges the
Kaluza-Klein states, carrying momenta in the internal
directions, with the magnetically charged soliton states in the
theory. Such a symmetry is necessarily non-perturbative, and
cannot be understandood from the point of view of
the string world-sheet theory, which is designed to produce the
perturbation expansion in string theory.

This distinction between the roles played by the SL(2,Z) and O(6,22;Z)
symmetries in string theory was already manifest in \S\ref{s2.2}, where we
saw that in the low energy effective field theory describing the four
dimensional string theory, the two symmetries appear on a somewhat
different footing. O(6,22;Z) is a symmetry of the effective action,
whereas SL(2,Z) is only a symmetry of the equations of motion.  However,
in  \S\ref{s2.4} we saw that with the restriction to field configurations
without any ten dimensional gauge fields, and by going to a dual formulation
of the theory, the roles of the SL(2,Z) and O(6,22;Z) symmetries can be
reversed. In this new formulation SL(2,Z) becomes a symmetry of the action,
whereas an O(6,6;Z) subgroup of the O(6,22; Z) group becomes a symmetry
only of the equations of motion.

This leads us to believe that if there is an alternate formulation of the
heterotic string theory, where the dual formulation of the $N=1$
supergravity theory in ten dimensions (or its dimensional reduction)
appears naturally as the low energy effective field theory in ten (or
four) dimensions, then SL(2,Z) transformations will have a more natural
action on the states in this new formulation.  Fortunately, it has already
been conjectured that such a dual formulation of the heterotic string
theory exists. It has been argued in Ref.\cite{DUFF} that heterotic string
theory is equivalent to a theory of 5 dimensional extended objects, also
known as 5-branes. The fields $\wt G^\ten_{MN}$, and $\wt
B^\ten_{M_1\ldots M_6}$, that appear in the dual formulation of the $N=1$
supergravity theory, have natural couplings to the five-brane.
(Unfortunately, at present there is no satisfactory way of coupling the
ten dimensional gauge fields to the five-brane, so we shall leave them out
of the analysis of this section. This difficulty may be related to the
difficulty that we encountered in \S\ref{s2.3} in writing down a
manifestly SL(2,R) and general coordinate invariant effective action in
the presence of ten dimensional gauge fields.)
Thus one might hope that the SL(2,Z) transformation has a natural action
on the five-brane world volume theory.

We shall now see that this is indeed the case\cite{SCHSEN2}. In
particular, we shall show that the quantum numbers $\alpha^m$ and
$\beta^m$ ($1\le m\le 6$) have interpretation as the internal momenta and
the five-brane winding numbers\cite{AZC}
along the internal direction respectively.
Thus the SL(2,Z) matrix $\pmatrix{0 & 1\cr -1 & 0}$, which corresponds to
the transformation $\alpha^m\to
\beta^m$, $\beta^m \to -\alpha^m$, exchanges the Kaluza-Klein
modes carrying internal momenta with the five-brane winding
modes on the torus. On the other hand, the quantum numbers
$\alpha^{m+6}$, $\beta^{m+6}$ ($1\le m \le 6$) correspond to
magnetic type charges in the five-brane theory, and only the
soliton solutions in the five-brane theory carry these charges.
As a result, part of the O(6,22;Z) symmetry,
$\alpha^m\leftrightarrow \alpha^{m+6}$, now interchanges
elementary excitations of the five-brane theory with the
solitons in this theory.

The world-volume swept out by the five-brane is six dimensional.
If $\xi^r$ denote the coordinates of this world volume ($0\le
r \le 5$) and $Z^M$ denote the coordinates of the ten
dimensional embedding space ($0\le Z\le 9$), then in the
presence of the background $\wt G^\ten_{MN}$ and $\wt
B^\ten_{M_1\ldots M_6}$,  the five-brane world-volume theory is described
by the action\cite{DUFFLUTWO}
\be \label{5b1}
\int d^6\xi \big[{1\over 2} \sqrt{-\gamma} \gamma^{rs} \wt
G^\ten_{MN} \p_r Z^M \p_s Z^N - 2\sqrt{-\gamma} + {1\over 6!}
\wt B^\ten_{M_1 \ldots M_6} \epsilon^{r_1 \ldots r_6} \p_{r_1}
Z^{M_1} \cdots \p_{r_6} Z^{M_6} \big].
\ee
Here $\gamma_{rs}$ is the metric on the five-brane world volume.  Upon
compactification, the coordinates $Z^M$ split into the space time
coordinates $X^\mu=Z^\mu$ ($0\le \mu \le 3$) and internal coordinates
$Y^m=Z^{m+3}$ ($1\le m\le 6$).  Let us first consider a background where
all fields are independent of the internal coordinates $Y^m$, and the only
non-vanishing components of the fields are
\be \label{5b2}
\wt G^\ten_{mn}, \quad \quad \wt G^\ten_{\mu\nu}, \quad \quad
\hbox{and} \quad \quad \wt B^\ten_{m_1 \ldots m_6} = \lambda_1
\epsilon_{m_1\ldots m_6}.
\ee
Furthermore, $\wt G^\ten_{\mu\nu}$ is adjusted so that
$g_{\mu\nu}=\eta_{\mu\nu}$ asymptotically.
The corresponding world-volume theory has two conserved current densities,
given by,
\ben \label{5b3}
j^r_m &=& \big( \sqrt{-\gamma}\, \gamma^{rs} \wt G^\ten_{mn} \p_s Y^n
+{\lambda_1\over 5!} \epsilon^{r r_2\ldots r_6} \epsilon_{m m_2
\ldots m_6} \p_{r_2} Y^{m_2}
\cdots \p_{r_6} Y^{m_6} \big) \nonumber \\
\wt j^r_m &=& {1\over 5!} \epsilon^{r r_2\ldots r_6} \epsilon_{m
m_2\ldots m_6} \p_{r_2} Y^{m_2}
\cdots \p_{r_6} Y^{m_6},
\een
which can be interpreted as the current densities associated with the
five-brane internal momenta and winding numbers respectively.
The total internal momenta $p_m$ and winding numbers $w_m$ of
the five-brane are given by,
\be \label{5b4}
p_m = \int d^5 \xi j^0_m, \quad \quad w_m = \int d^5\xi \wt
j^0_m.
\ee

In order to find the relationship between these conserved charges,
and the quantum
numbers $\alpha^m$ and $\beta^m$, we shall proceed in three
stages. In the first stage we shall determine the coupling of
the background gauge fields $\C^m_\mu$ and $\D^m_\mu$, defined
through Eq.\refb{1.26}, to the current densities $j^r_m$ and $\wt j^r_m$.
In the second stage, we shall calculate the asymptotic values of the
fields $F^{(\C)m}_{\mu\nu}$ and $F^{(\D)m}_{\mu\nu}$ in the presence of a
five-brane carrying a fixed amount of $p_m$ and $w_m$ charges. In the
third stage, we shall relate the asymptotic values of $F^{(\C)m}_{\mu\nu}$
and $F^{(\D)m}_{\mu\nu}$ to the asymptotic values of $F^\al_{\mu\nu}$, and
hence to $\alpha^a$ and $\beta^a$.

In order to carry out the first step, we switch on the background
fields $\wt G^\ten_{m\mu}$ and $\wt B^\ten_{\mu m_2\ldots m_6}$,
and calculate the resulting contribution to the five-brane world
volume action to linear order in these fields. Using
Eqs.\refb{5b1} and \refb{1.26} we find that the extra
contribution to the  action to linear order in $\C^m_\mu$ and
$\D^m_\mu$ is given by
\be \label{5b5}
\int d^6\xi (\C^m_\mu j^r_m \p_r X^\mu + \D^m_\mu \wt j^r_m \p_r
X^\mu).
\ee
Using the identification \refb{1.35}, we can rewrite this
coupling as
\be \label{5b6}
2 \int d^6 \xi (\wc A^{(m,1)}_\mu j^r_m \p_r X^\mu - \wc
A^{(m,2)}_\mu \wt j^r_m \p_r X^\mu).
\ee
If we work in the static gauge $\xi^0=X^0$, then the coupling of
$\wc A^{(m,\alpha)}_0$ is given by,
\be \label{5b6new}
2 \int d^5 \xi d X^0 (\wc A^{(m,1)}_0 j^0_m  - \wc A^{(m,2)}_0 \wt
j^0_m ).
\ee

We now add \refb{5b6new} to the action \refb{1.21} (or, equivalently,
\refb{1.33}), derive the equations of motion for the gauge fields $\wc
A^{(m,\alpha)}_\mu$, and compute the fields $\wc F^{(m,\alpha)}_{\mu\nu}$
induced by the 5-brane source. The resulting asymptotic values of these
fields are given by the equations
\be \label{5b7}
\wh G^\zero_{mn} (\LL^T \M^\zero \LL)\pmatrix{\wc F^{(n,1)}_{0r} \cr
\wc F^{(n,2)}_{0r} } + \wh B^\zero_{mn} \LL
\pmatrix{\wc{\tF}^{(n,1)}_{0r} \cr
\wc{\tF}^{(n,2)}_{0r} } \simeq {4\over r^2} \pmatrix{ - \int d^5 \xi j^0_m
\cr \int d^5 \xi \wt j^0_m} = {4\over r^2} \pmatrix{ - p_m \cr w_m},
\ee
and
\be \label{5b7dual}
\pmatrix{\wc{\tF}^{(m,1)}_{0r} \cr
\wc{\tF}^{(m,2)}_{0r} }\simeq 0.
\ee
This determines the asymptotic values of the fields
$\wc F^{(m,\alpha)}_{\mu\nu}$ for $1\le m\le 6$ and $1\le \alpha \le
6$. On the other hand, the quantum numbers $\alpha^a$ and $\beta^a$
are related to the asymptotic values of the fields
$\wc F^\al_{\mu\nu}$ for $1\le a\le 12$, as can be seen from
Eqs.\refb{defqea}, \refb{defqma} and \refb{solitonexit}. In the
source free region, the relationship between the two sets of
fields $\wc F^{(m, \alpha)}_{\mu\nu}$ and $\wc F^\al_{\mu\nu}$ can be
found by starting with the action \refb{1.20a}, writing down
the gauge field equations of motion in this theory, and noting that $\wc
F^\al_{\mu\nu} \equiv \wc F^{(a,1)}_{\mu\nu}$ for $1\le a\le 12$. These
equations let us express $\wc F^\al_{\mu\nu}$ in terms of the
fields $\wc F^{(m,\alpha)}_{\mu\nu}$, from which we can calculate
the asymptotic values of the fields $\wc F^\al_{\mu\nu}$ in terms of
$p_m$ and $w_m$. Comparing these asymptotic values with Eqs.\refb{defqea},
\refb{defqma} we get,
\ben\label{newqm}
Q^{(m)}_{el} = {4\over \lambda_2^\zero} \wh G^{\zero mn} (-p_n +
\lambda_1^\zero w_n), \quad && \quad Q^{(m)}_{mag} =0 \nonumber \\
Q^{(m+6)}_{el} = - \, {4\over \lambda_2^\zero}
\wh B^\zero_{mq} \wh G^{\zero qn}(-p_n
+ \lambda_1^\zero w_n), \quad && \quad Q^{(m+6)}_{mag} =4 w_m.
\een
(Note that when $A^{\ten I}_M=0$, then $\wc
F^\al_{\mu\nu} = F^\al_{\mu\nu}$ for $1\le a\le 12$.)
Finally, comparison with Eq.\refb{solitonexit} yields
\be
\alpha^m = - 4 p_m, \quad\quad \beta^m = 4 w_m, \quad \quad
\alpha^{m+6} =\beta^{m+6} =0, \quad\quad \hbox{for}\quad 1\le
m\le 6.
\ee
(Note that here $\vec\alpha$ and $\vec\beta$ are 12 dimensional
vectors, since we have ignored the charges associated with the
ten dimensional gauge fields.) This establishes the desired
relation, i.e. the quantum numbers $\alpha^m$ and $\beta^m$ are
related to the five-brane momenta and winding numbers in the
internal direction respectively. Thus we see that the SL(2,Z)
transformations do interchange the Kaluza-Klein modes with the
five-brane winding modes.\footnote{ This conclusion is also
consistent with the fact that the $H$-monopole solutions can be
regarded as five branes wrapped around the torus\cite{GHL}.}
Note also that the quantum numbers $\alpha^{m+6}$ and
$\beta^{m+6}$ for $1\le m\le 6$ now have to be interpreted as
topological charges in the five-brane theory.

There are in fact further analogies between the target space
duality transformations in string theory and the SL(2,Z)
transformations in the five-brane theory. Let us define
\be\label{new5one}
\G_{mn} = \wt G^\ten_{m+3, n+3}
\ee
as the internal components of the five-brane metric. {}From
Eqs.\refb{1.26} we see that the complex field $\lambda$ has a
natural expression in terms of the variables in the five-brane
theory:
\be \label{new5two}
\lambda = \wt B^\ten_{4\ldots 9} + i\sqrt{\det \G}.
\ee
This is very similar to the expression for the complex structure
moduli field $\tau$ for string theory compactified on a two
dimensional torus:
\be \label{new5three}
\tau = B^\ten_{89} + i \sqrt{\det \bar G},
\ee
where 8 and 9 denote the compact directions and $\bar G$ denote
the components of $G^\ten$ in the two internal directions. Here
$B^\ten$ and $G^\ten$ are the variables that couple naturally to
the string.
Under the target space duality transformation, the variable
$\tau$ transforms to $(a\tau + b)/(c\tau +d)$ with $\pmatrix{a &
b\cr c & d}$ an SL(2,Z) matrix, exactly as $\lambda$ transforms
under the S-duality transformation.

The existence of target space duality symmetry in string theory implies
the existence of a minimum compactification radius, since the
T-duality transformation relates tori of small radius to tori of
large radius, with distances measured in the string metric $G^\ten_{MN}$.
In the same spirit, the S-duality symmetry in string theory implies the
existence of a maximum value of the string coupling constant. The
discussion in the previous paragraph shows that this result may also be
interpreted as the existence of a minimum size of the compact manifold,
but now measured in the five-brane metric $\wt G^\ten_{MN}$.

We end this section by summarising the roles of SL(2,Z) and
O(6,6;Z) transformations in the string theory and the five-brane
theory. This is best illustrated in the following table:

\begin{center}
\begin{tabular}{|| l | l ||}
\hline & \\ \hline
{}~ \quad \quad String Theory & {}~ \quad \quad Five Brane Theory \\
\hline  & \\ \hline
O(6,6;Z) is the symmetry of the &  SL(2,Z) is the symmetry of the \\
low energy effective action  &  low energy effective action \\
\hline
SL(2,Z) is the symmetry of the   &  O(6,6;Z) is the symmetry of the \\
low energy equations of motion     &  low energy equations of
motion \\
\hline
O(6,6;Z) exchanges Kaluza-Klein & SL(2,Z) exchanges
Kaluza-Klein \\
modes with string winding modes & modes with 5-brane winding
modes \\
\hline
SL(2,Z) exchanges elementary  & O(6,6;Z) exchanges elementary \\
string excitation with solitons & 5-brane excitations with
solitons \\
in string theory  & in 5-brane theory \\
\hline
O(6,6;Z) implies a minimum size of  & SL(2,Z) implies a minimum
size of \\
the compact manifold measured in  & the compact manifold measured in \\
the string metric   &  the 5-brane metric \\
\hline & \\ \hline
\end{tabular}
\end{center}

\section{Discussion and Open Problems}\label{s6}

We conclude these notes with a discussion of some specific features of the
SL(2,Z) symmetry, and some open problems in this area.

\subsection{SL(2,Z) as a Discrete Gauge Symmetry}\label{dis1}

We have already argued that S-duality transformation has the possibility
of being a symmetry of the four dimensional heterotic string theory.  We
shall now show that if SL(2,Z) is a symmetry of the theory, then it must
act as a discrete gauge symmetry, {\it i.e.} we must {\it identify} field
configurations that are related by any SL(2,Z) transformation. To start
with, we note that the full SL(2,Z) group is generated by two elements,
\be\label{SL2Zelem}
\T = \pmatrix{1 & 1\cr 0 & 1}, \quad \quad \SS = \pmatrix{0 & 1\cr -1 & 0}.
\ee
$\T$ generates the transformation $\lambda\to \lambda+1$. It is well
known\cite{WITTENSTRING,DABHOLKAR} that $\lambda$ changes by 1 as we go
around an elementary string. As a result, the very existence of elementary
string states forces us to identify field configurations related by the
transformation $\T$. Now, if $\SS$ is a symmetry of the theory, then,
acting on an elementary string state it must produce a valid state in the
theory. But when we go around this new state, the field configuration
changes by the SL(2,Z) transformation $\SS\T\SS^{-1}$. Thus we must also
identify field configurations that are related by the SL(2,Z)
transformation $\SS\T\SS^{-1}$. Now, Eq.\refb{SL2Zelem} gives
\be\label{texpress}
\SS= \T\cdot \SS\T\SS^{-1}\cdot \T,
\ee
showing that the full SL(2,Z) group is generated by $\T$ and
$\SS\T\SS^{-1}$.  This shows that we must identify field configurations
which are related by any SL(2,Z) transformation, {\it i.e.} SL(2,Z) must
be treated as a discrete gauge symmetry of the theory.

\subsection{Relation to Other Proposals}

Electric-Magnetic duality in four dimensional string theory has been
discussed from a different point of view in Refs.\cite{DUFFEM}. This
duality transformation can be identified to the string - five-brane
duality transformation, when both the string theory and the five-brane
theory are compactified on a six dimensional torus. This differs from
the duality symmetry discussed here in an essential way, namely the
string - five-brane duality transformation relates two
different theories, and in that sense, is not a {\it symmetry} of
any theory, whereas the SL(2,Z) transformation discussed here relates
two different vacua of the same theory. This can also be seen
from the point of view of the low energy effective field theory,
$-$ SL(2,Z) acts as a transformation on the variables of the low
energy effective field theory, and is a symmetry of the
equations of motion in the theory, whereas the string -
five-brane duality transformation relates variables of two
different actions \refb{1.20} and \refb{1.33}.

\subsection{Open Problems}

In this paper we have produced several pieces of evidence for the
existence of SL(2,Z) symmetry in string theory compactified on a six
dimensional torus. However, much work remains to be done. First of all, we
need to explicitly construct the new monopole and dyon states in the
theory which must exist in order for SL(2,Z) to be a genuine symmetry.
These have been discussed in \S\ref{snew}, but we shall list them again
here.

1) SL(2,Z) symmetry predicts the existence of BPS dyon solutions (with
space-like electric and magnetic charge vectors) carrying multiple units
of magnetic and electric charge in the vector representation of the N=4
super-Poincare algebra. Furthermore, if $p$ and $r$ denote the number of
units of electric and magnetic charges carried by the dyon, then $p$ and
$r$ must be relatively prime. For $r>1$, these dyons could be regarded as
supersymmetric bound states of monopoles and dyons, each carrying single
unit of magnetic charge.  A careful quantization of the zero modes of the
BPS multi-monopole solutions\cite{CORRIGAN} should exhibit these features
if SL(2,Z) is a genuine symmetry of the theory. Recent results of
Ref.\cite{BLUM}, as well as earlier results of
Refs.\cite{GIBMAN,HARSTR,GAUNT} may be particularly useful for this
purpose. Triangle inequality guarantees that the energy of a
supersymmetric state carrying these charges is strictly less than the
lowest energy state in the continuum, hence it is quite plausible that
such bound states do exist in the theory.

2) SL(2,Z) symmetry also predicts the existence of $H$-monopole and dyon
solutions (with light-like electric and magnetic charge vectors) carrying
multiple units of electric and magnetic charge. As before, if $p$ and
$r$ denote the number of units of electric and magnetic charge carried by
the dyon, then $p$ and $r$ must be relatively prime. For each such pair
$(p,r)$ there should be 21 distinct dyon states in the vector
supermultiplet of the N=4 super-Poincare algebra, and one dyon state in
the spin 2 representation of the N=4 super-Poincare algebra, saturating
the Bogomol'nyi bound. Finally these solutions must exist at any generic
point in the compactification moduli space. At present the existence of
such solutions has been shown only at special points in the moduli space,
where there is one or more unbroken SU(2) gauge group.

3) Finally, SL(2,Z) symmetry predicts the existence of monopole and
dyon solutions with time-like electric and magnetic charge vectors.
However, there is no limit in which these states become massless. As a
result we do not expect these states to be represented as solutions in the
effective field theory involving (nearly) massless fields. Perhaps one
might be able to construct them as exact conformal field theories.

Another useful direction of investigation may be the study of five-branes.
We have argued that the SL(2,Z) transformations act naturally on the
five-branes, and hence it might be possible to establish that
the five brane theory has an exact SL(2,Z) symmetry, even if we cannot
solve the five-brane theory. This would at least establish that the
SL(2,Z) symmetry of the four dimensional string theory is an immediate
consequence of the string$-$five-brane duality in arbitrary dimensions.

{\bf Acknowledgements}: I would like to thank J. Schwarz for collaboration
in Refs.\cite{SCHSEN1,SCHSEN2} and many discussions. I would also like to
thank M. Duff, A. Strominger and P. Townsend for discussions.  Finally I
would like to thank the Institute of Physics, Bhubaneswar for hospitality
during preparation of this manuscript, and S. Rao for a critical reading
of the manuscript.

\end{document}